\documentclass[prx,twocolumn,showpacs,superscriptaddress]{revtex4-1}
\usepackage[english]{babel}
\usepackage[dvips]{graphicx}
\usepackage{color}
\usepackage{latexsym}
\usepackage{amsmath}
\usepackage{amsthm}
\usepackage{amsfonts}
\usepackage{amssymb}
\usepackage{mathtools}
\usepackage{cancel}

\begin{document}

\pagenumbering{arabic}

\title{Effects of a dissipative coupling to the momentum of a particle in a double well potential}
%\title{Dissipative effects in a double well potential}
%\title{Dissipative momentum coupling effects in a double well potential}
%\title{Momentum dissipation effects in a double well potential}
%
%
%
\author{D.\ Maile}
\affiliation{Fachbereich Physik, Universit{\"a}t Konstanz, D-78457 Konstanz, Germany}
\affiliation{Institut f\"ur Theoretische Physik and Center for Quantum Science, Universit{\"a}t T{\"u}bingen, Auf der Morgenstelle 14, 72076 T{\"u}bingen, Germany}
\author{S. Andergassen}
\affiliation{Institut f\"ur Theoretische Physik and Center for Quantum Science, Universit{\"a}t T{\"u}bingen, Auf der Morgenstelle 14, 72076 T{\"u}bingen, Germany}
\author{G. Rastelli}
\affiliation{Fachbereich Physik, Universit{\"a}t Konstanz, D-78457 Konstanz, Germany}
\affiliation{Zukunftskolleg, Universit{\"a}t Konstanz, D-78457, Konstanz, Germany}
\begin{abstract}
Double well potentials offer the possibility of coherent state preparation and therefore constitute important building blocks in the analysis of quantum information and quantum engineering devices. Here we present a study of the coherent tunneling in a parabolic double well potential in presence of different dissipative interactions. Specifically, we investigate the effects of an environmental coupling to the momentum and/or to the position of a particle in the potential. Using the semiclassical approximation to calculate instanton paths in Euclidean time, we find that momentum dissipation enhances the coherent tunnel splitting.  In presence of both types of dissipation, the momentum dissipation shifts the critical coupling strength of the dissipative phase transition induced by the position dissipation.  

%
%Such a model can be realized in Josephson junction chains with shunt resistances and resistances between 
%the chain and the ground. 
%
%Using a self-consistent harmonic approximation, we determine the phase diagram at zero temperature which exhibits 
%a quantum phase transition between an ordered phase, corresponding to the superconducting state, 
%and a disordered phase, corresponding to the insulating state with localized superconducting charge. 
%
%nterestingly, we find that the critical line separating the two phases has a non monotonic behavior as a function of the dissipative 
%coupling strength. 
%
%This result is a consequence of the frustration between (i) one dissipative coupling that quenches the quantum phase fluctuations 
%favoring the ordered phase and (ii) one that quenches the quantum momentum (charge) fluctuations leading to a vanishing phase coherence.
%
%Moreover, within the self-consistent harmonic approximation, we analyze the dissipation induced crossover between a first and second order phase transition, showing that quantum frustration increases the range in which the phase transition is second order.
%
%The non monotonic behavior is reflected also in the purity of the system that quantifies the degree of correlation between the system and the environment, 
%and in the logarithmic negativity as entanglement measure that encodes the internal quantum correlations in the chain.
%
\end{abstract}
\date{\today}
\maketitle

%%%%%%%%%%%%%%%%%%%%%%%%%%%%%%%%%%%%%%%%%%%%%%%%%%%%%%%%%%%%%%%
%%%%%%%%%%%%%%%%%%%%%%%%%%%%%%%%%%%%%%%%%%%%%%%%%%%%%%%%%%%%%%%
%%%
%%%
%%%	SEC.1   INTRODUCTION
%%%
%%%
%%%
%%%%%%%%%%%%%%%%%%%%%%%%%%%%%%%%%%%%%%%%%%%%%%%%%%%%%%%%%%%%%%%
%%%%%%%%%%%%%%%%%%%%%%%%%%%%%%%%%%%%%%%%%%%%%%%%%%%%%%%%%%%%%%%
%
%
%INTRODUCITION%%%%%%%%%%%%%%%%%%%%%%%%%%%%%%%%%%%%%%%%%%

\section{Introduction}
Preserving and controlling quantum behavior 
is of main interest in today's research, paving the way to quantum engineered systems like quantum computers and quantum simulators \cite{Nielsen:2010}. 
In experimental setups the coupling to environmental degrees of freedom is one of the main sources of decoherence. To theoretically describe these systems accurately we therefore have to take unavoidable effects of the environment into account \cite{Weiss:2012,BRE02}. 
On the other hand, recent progress in  experimental setups promise the realizability of reservoir engineered quantum systems, aiming to simulate dissipative quantum problems or to control quantum information systems \cite{Verstaete:2009,Mirrahimi:2014,Poyatos:1996,Shankar:2013,Ankerhold:2019}. 
In these perspectives, the seminal theoretical work of  Caldeira and Leggett on open quantum systems is still of major importance in today's research. They studied dissipative effects on quantum tunneling, in particular the decay out of metastable potentials affected by dissipation, by using a phenomenological model, in which the environment consists of a bath of harmonic oscillators with a given spectral density \cite{Caldeira:1981tun,Caldeira:1982ann}.
Since then the Caldeira-Leggett model had a huge influence on the theory of open quantum systems with many applications \cite{Weiss:2012,BRE02}.

Double well potentials, potentially describing a qubit, are important in the context of dissipative effects on quantum information systems.
In the reduction to a two level system, the analysis of quantum dissipation was investigated using  the spin - boson model \cite{Leggett:1987,Dorsey:1986}.
In extensions beyond this reduced scheme, the particle in the full potential is coupled to a Caldeira-Leggett bath of harmonic oscillators via its position coordinate (conventional dissipation). While early works studied the second problem in the ohmic regime analytically using the instanton technique \cite{Bray:1982,Chakravarty:1982,Grabert:1987}, other groups used numerical Monte Carlo or renormalization group techniques   \cite{Matsuo1:2004,Matsuo2:2008,Aoki:2002, Kovacs:2017}.
All these approaches yield the important result that dissipation destroys the coherent superposition states leading to a classical localized state (dissipative phase transition). 
Contrarily, Fujikawa et al. observed an enhancement of quantum tunneling in the double well by dissipative interactions in the super-ohmic regime,  because of a virtual mixing of the ground states with excited states \cite{Fujikawa:1992}. 
In the limit of low temperatures, when the particle is coupled to an Ohmic environment through a given operator, the quantum fluctuations of the respective observable are squeezed. This effect leads to an enhancement of the quantum fluctuations of canonically conjugate observables, since the Heisenberg uncertainty has to be fulfilled.
For example, coupling the environment to the momentum of the particle increases the quantum fluctuations of the position.
Such a dissipative mechanism can lead to an enhancement of quantum effects \cite{Ankerhold:2007}.
When both dissipative couplings are present (termed dissipative frustration), the quantum fluctuations in a harmonic oscillator become non monotonic as a function of the dissipative coupling  \cite{Cuccoli:2010dr,Rastelli:2016ge,Kohler:2006ky}.
Dissipative momentum and position couplings can be translated to dissipative charge  or dissipative phase couplings in electrical quantum circuits, respectively \cite{Devoret:2004-LesHouches}.  In these systems a variety of effects, including dissipative quantum phase transitions have been observed \cite{Fazio:2001,Baranger:2014,Otten:2017}.
In the frustrated case in presence of both phase and charge dissipation, the phase diagram of a superconducting chain shows a non monotonic behavior as function of the dissipative coupling \cite{Maile:2018}. 
On the other hand, coupling two bosonic baths within the spin-boson model to two non commuting observables in the ohmic regime (e.g $\sigma_x$ \textit{and} $\sigma_y$) results in a canceling effect of both couplings  preserving quantum tunneling \cite{Bruognolo:2014jf,Neto:2003ir}. 

In this work we study how momentum (which we denote as unconventional) dissipation affects the tunnel splitting of a particle in a parabolic double well potential.
We investigate this system in the limit of zero temperature, where quantum fluctuations  are  dominant. Further, we  analyze the interplay with conventional position dissipation and its implications on the dissipative phase transition.
We use the Euclidean path integral method to calculate the diagonal density matrix element in presence of both dissipative couplings in the semiclassical limit.
The instanton method \cite{,Coleman:1978ae,Kleinert:2009} allows us to evaluate the tunnel splitting. Employing Anderson's renormalization group approach \cite{Anderson:1970}, we furthermore determine its effects on the phase transition.
We find that the tunnel splitting is enhanced by increasing the dissipative momentum coupling. 
For the case of both dissipative couplings present, the momentum dissipation renormalizes the critical coupling strength for the phase transition induced by the position dissipative interaction.
Specifically, we find that larger values of the coupling strength are necessary in order to yield the particle localization.
This result is consistent with the interpretation that the momentum dissipation increases the position quantum fluctuations and, hence,  competes with the mechanism of classical localization of the particle caused by the position dissipation. 
The effects we discuss have their origin in the Heisenberg uncertainty relation between the momentum $\hat{p}$ and the position $\hat{x}$. Our system is therefore fundamentally different from the spin-boson model. 

The paper is organized as follows. In Section~\ref{Sec.Model} we introduce the parabolic double well model together with the ohmic dissipative couplings to the position and momentum. %we take into account. 
We discuss its treatment in the semiclassical approximation within the instanton technique, which we generalize to the case of momentum dissipation.
%Then we use the established mapping to a one dimensional gas of interacting charges  \cite{Chakravarty:1982}. 
In Section~\ref{Sec.results} we present our results for the different dissipative cases together with the basic steps of the analytic calculations. For more details we refer to the Appendices. We draw our conclusions in Section~\ref{Sec.Conclusion}. 

%%%%%%%%%%%%%%%%%%%%%%%%%%%%%%%%%%%%%%%%

%MODEL AND APPROXIMATION%%%%%%%%%%%%%%%%%%%%%%%%%%%%%%%%

\section{Model and Approximations}
\label{Sec.Model}

In this section we introduce the parabolic double well, the dissipative interactions, the semiclassical instanton method and discuss the mapping of the dissipative tunneling problem to a one dimensional gas of interacting charges.

%Fig.1%%%%%%%%%%%%%%%%%%%%%%%%%%%%%%%%%%%

\begin{figure}[t]
	\centering
	\includegraphics[width=1.00\linewidth]{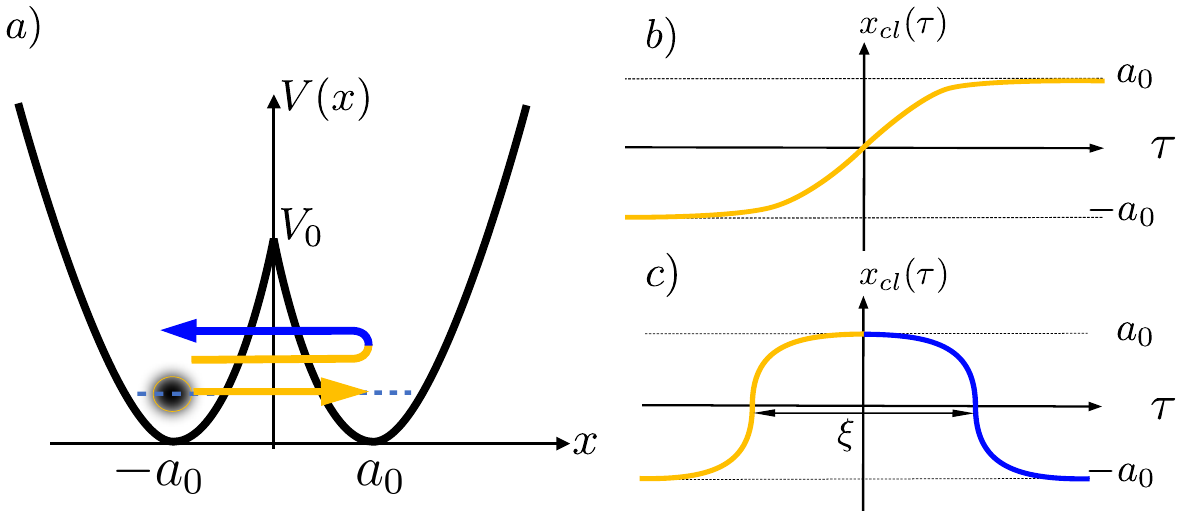}
	\caption{a) Parabolic double well potential with a barrier height  $V_0=\frac{1}{2}m\omega_0^2 a_0^2$ and minima in $\pm a_0$.  b) Single instanton path. c) Bounce path of distance $\xi$ formed by two instantons of opposite sign. }
	\label{Fig.Instanton}
\end{figure}
%%%%%%%%%%%%%%%%%%%%%%%%%%%%%%%%%%%%%%%

\subsection{The action of the parabolic double well in presence of conventional and unconventional dissipation}

We consider the symmetric parabolic double well potential
\begin{align}
V(x)=\frac{m\omega_{0}^{2}}{2}\left(\theta(-x)(x+a_{0})^{2}+\theta(x)(x-a_{0})^{2}\right)\;,\label{Eq.potential}
\end{align}
with $m$ the mass of the particle, $\pm a_0$  the positions of the minima and $\omega_0$ the characteristic frequency of each parabola (see Fig.~\ref{Fig.Instanton}a). By construction the barrier height between the wells is $V_0=\frac{1}{2}m\omega_0^2a_0^2$.  

We introduce the dissipation via the Caldeira-Legget harmonic oscillator baths. The  starting  Hamiltonian reads
\begin{align}
\mathcal{H}=\mathcal{H}_{sys}+\mathcal{H}_{B,\hat{x}}+\mathcal{H}_{B,\hat{p}}\;,
\end{align}
where the part for the bare system is $\mathcal{H}_{sys}=\hat{p}^2/2m+V(\hat{x})$ with $\hat{x}$ and $\hat{p}$  the position and  the momentum operator of the particle, respectively. The second part 
\begin{align}
\mathcal{H}_{B,\hat{x}}=\frac{1}{2}\sum_i\left[\frac{\hat{P}_i}{M_i}+M_i\omega_i^2\left(\hat{X}_i-\frac{\lambda_i}{M_i\omega_i^2}\hat{x}\right)^2\right]\label{Eq.CLx}
\end{align}
 is the bath coupled to the position with $\hat{X}_i$ the position and $\hat{P}_i$ the momentum operator of the respective harmonic oscillator, while
\begin{align}
\mathcal{H}_{B,\hat{p}}=\frac{1}{2}\sum_j\left[\frac{\left(\hat{P}'_j-\mu_j\hat{p}\right)^2}{M_j}+M_j\omega_j^2\hat{X}'^2_j\right]\label{Eq.CLp}
\end{align}
is the second bath coupled to the momentum, again with $\hat{X}'_j$ and $\hat{P}'_j$ being the position and the momentum operator of the respective harmonic oscillator $j$.  

We use the imaginary time path integral formalism to calculate the density matrix elements of the system \cite{Weiss:2012}.  We integrate out the momentum $\hat{p}$, the bath coordinates $\hat{P}_i, \hat{X}_i$ and $\hat{P}'_j, \hat{X}'_j$ leading to an effective Euclidean action of the system
\begin{align}
S[x]&=\int_{-\frac{\beta}{2}}^{\frac{\beta}{2}}d\tau\left(\frac{m}{2}\dot{x}^{2}(\tau)+V[x(\tau)]\right)+S_{dis}[x]\;,
\label{Eq. action1}
\end{align}
with imaginary time $\tau$, $\beta=\hbar/k_BT $ is related to the inverse temperature and $S_{dis}$ containing the dissipative effects. In the calculation of the dissipative part of the action we chose ohmic spectral densities for the environments \cite{Maile:2018}. This yields 
\begin{align}
S_{dis}&=\frac{1}{2}\iint_{-\frac{\beta}{2}}^{\frac{\beta}{2}}d\tau d\tau'x(\tau)x(\tau')F(\tau-\tau')\nonumber\\
&+\frac{1}{2}\iint_{-\frac{\beta}{2}}^{\frac{\beta}{2}}d\tau d\tau'\dot{x}(\tau)\dot{x}(\tau')\widetilde{F}(\tau-\tau')\;, \label{Eq.dissaction}
\end{align}
where $F(\tau)=\frac{m}{\beta}\sum_{l=-\infty}^{\infty}\gamma {|\omega_{l}|}e^{i\omega_l\tau}$ is the ohmic kernel for the conventional position dissipation and $\widetilde{F}(\tau)=\frac{m}{\beta}\sum_{l=-\infty}^{\infty}\frac{-\tau_{p}|\omega_{l}|}{1+\tau_{p}|\omega_{l}|f_c\left(\frac{|\omega_l|}{\omega_c} \right)}e^{i\omega_l\tau}$ the one for the unconventional momentum dissipation. The position fluctuations diverge in presence of momentum dissipation. This divergence is cured by the Drude cutoff function $f_c\left(|\omega_l|/\omega_c\right)=\left(1+{|\omega_l|}/{\omega_c}\right)^{-1}$, where $\omega_c$ is the high frequency cutoff \cite{Weiss:2012}. $\gamma$ and $\tau_p$ 
are the coupling parameters to the respective baths and $\omega_l=2\pi l/\beta$ are Matsubara frequencies. Note  that although Eqs.~(\ref{Eq.CLx}) and (\ref{Eq.CLp}) are symmetric for the coupling in position and momentum in the Hamiltonian representation, while Eq.~(\ref{Eq.dissaction}) is not symmetric anymore, as we use the path integral in position representation. 
Finally, the unnormalized density matrix elements of the effective system given by the particle in the double well affected by dissipation reads
\begin{equation}
\rho_{x_1,x_2}=\int_{\substack{ \\\substack{x(-\frac{\beta}{2})=x_1 \\x(+\frac{\beta}{2})=x_2}}}\mathcal{D}[x(\tau)]\;e^{-\frac{1}{\hbar}S[x]}\label{Eq.density1}\;, 
\end{equation}
where the integral extends over all possible paths starting in $x_1$ and ending at $x_2$. 

\subsection{Semiclassical approximation}
We restrict our discussion to the so-called semiclassical limit in imaginary time, where $V_0\gg\hbar\omega_0$.
Here the classical path in the path integral Eq.~(\ref{Eq.density1}) yields the largest contribution to the density matrix element. 
Using such a path, we expand $x(\tau)\approx x_{cl}(\tau)+\delta x(\tau)$, where $x_{cl}(\tau)$ is the path that minimizes the action in Eq.~(\ref{Eq. action1}) with the boundary conditions $x_{cl}(-\beta/2)=x_1$ and $x_{cl}(\beta/2)=x_2$. The part $\delta x(\tau)$ represents the fluctuations around this path satisfying $\delta x(\pm \beta/2)=0$.
The minimizing path is also known as "classical path" since it can be mathematically obtained as the solution of a Newton equation (non local in time) of the particle moving in the inverted potential $-V(x)$  and in the presence of dissipation. However, it does not represent at all the classical physical path of the particle moving in the double-well potential. The latter path can not exist classically since the particle has to be confined in one of the two wells in the ground state. Within this expansion, we find for the action Eq.~(\ref{Eq. action1}) $S[x]\approx S[x_{cl}(\tau)]+S_\delta[\delta x]$, where the first part is the action Eq.~(\ref{Eq. action1}) on the given path $ x_{cl}(\tau)$ and the second %part
\begin{align}
S_\delta[\delta x]&=\int_{-\frac{\beta}{2}}^{\frac{\beta}{2}}d\tau\left[\frac{m}{2}\delta\dot{x}^{2}(\tau)+\frac{1}{2}\frac{d^{2}V[x(\tau)]}{dx^{2}} \bigg|_{x_{cl}(\tau)} \delta x^{2}(\tau)\right]\nonumber \\
&+S_{dis}[\delta x] \label{Eq.deltaaction}
\end{align}
contains the fluctuations.
In the following discussion we are interested in the diagonal elements $\rho_{\pm a_0,\pm a_0}$  of the density matrix of the particle. These are given by the sum of all paths starting at $x_1=-a_0$ and ending at $x_2=-a_0$. In the semiclassical limit, we consider all possible paths that minimize the action as well as the fluctuations around them. These extreme paths are given by the sum of $n$ individual paths that cross the origin $x=0$ once. Such a path with a single crossing is called instanton $x^{(1)}_{cl}(\tau)$ (see Fig.~\ref{Fig.Instanton}b). For example, two instantons form a bounce path $x_{cl}^{(2)}$ with spacing $\xi$, as shown in Fig.~\ref{Fig.Instanton}c. The diagonal elements $\rho_{a_0,a_0}$ and $\rho_{-a_0,-a_0}$ include only paths formed by an even number of instantons
\begin{align}
\rho_{-a_0,-a_0}&=\sum_{n \text{ even}} \rho_{-a_0,-a_0}^{(n)}\;, \label{Eq.Sumrho}
%\rho_{-a_0,a_0}&=\sum_{n \text{ odd}} \rho_{-a_0,a_0}^{(n)}\;.
\end{align}
where each addend can be written in the form 
\begin{align}
\rho^{(n)}_{-a_0,-a_0}=\mathcal{F}^{(n)}e^{-\frac{1}{\hbar}S[x^{(n)}_{cl}]}\;,
\end{align}
where 
\begin{align}
\mathcal{F}^{(n)}=\oint_{\substack{\delta x(\frac{\beta}{2})=0\\ \delta x{(-\frac{\beta}{2})}=0}}\mathcal{D}[\delta x(\tau)]\;e^{-\frac{1}{\hbar}S_{\delta,n}[\delta x]} \label{Eq.prefactornodis}
\end{align}
accounts for the quantum fluctuations around the classical path.

\subsection{The instanton technique}
The path of the first contribution ($n=1$) is the single instanton path shown in Fig.~\ref{Fig.Instanton}b. The action on this classical path diverges in presence of conventional dissipation. Therefore, we cannot calculate the action for one instanton. To make the action convergent we introduce a second instanton at a distance $\xi$, as shown in Fig.~\ref{Fig.Instanton}c.  With this bounce path contribution we calculate $\rho_{-a_0,-a_0}^{(2)}=z_L^{(2)}$ of the diagonal density matrix element. More generally, we can calculate the contribution formed by a given number of instanton pairs with opposite direction (bounces). In this way we can write 
\begin{align}
\rho_{-a_{0},-a_{0}}	= Z_L=\sum_{k=0}^{\infty} z_L^{(k)},
\end{align}
where $n=2k$ single  instantons and $z_L^{(k)}=\rho_{-a_{0},-a_{0}}^{(2k)}$.

\subsubsection{The action on the instanton bounce path}
In order to calculate the action on the classical path, we make the ansatz for the classical bounce ($n=2$) starting at $x_1=-a_0$ and ending at $x_2=-a_0$ %reading
\begin{align}
x^{(2)}_{cl}(\tau)={x_{T}}+\sum_{l=1}^{\infty}\frac{v_{l}}{\omega_{l}}\left[\sin(\omega_{l}(\tau+\frac{\xi}{2}))-\sin(\omega_{l}(\tau-\frac{\xi}{2}))\right]\;,\label{Eq.bounce}
\end{align}
where $x_T=-a_0+v_{0}\xi$ and $v_0=2a_0/{\beta}$. This ansatz gives $x(\tau=0)\rightarrow a_0$ for $\xi\rightarrow \infty$ and has been obtained by combining an instanton $x_{cl}^{(1)}(\tau)$ going from $-a_0$ to $a_0$ with an instanton $-x_{cl}^{(1)}(\tau)$ for the opposite event centered at different times. Inserting the ansatz into Eq.~(\ref{Eq. action1}) and minimizing the action with respect to $v_l$, we find \footnote{Note that the action on the classical path is convergent and  the effect of the high frequency cutoff function $f_c(|\omega_l|/\omega_c)$ is irrelevant here.}
\begin{equation}
v^{(cl)}_{l}=\frac{4\omega_{0}^{2}a_{0}}{\beta\left(\omega_{0}^{2}+\frac{\omega_{l}^{2}}{1+\tau_{p}\omega_{l}}+\gamma\omega_{l}\right)}\;.\label{Eq.velocityclasscialpathmatsu}
\end{equation}
We insert this expression for $v_l^{(cl)}$ into the action and obtain
\begin{align}
S[x_{cl}^{(2)}]&=\frac{-16V_0}{\beta}\sum_{l=1}^{\infty}\frac{\omega_0^2}{\omega_{l}^{2}}\frac{\left(1-\cos\left(\omega_{l}\xi\right)\right)}{\left(\omega_{0}^{2}+\frac{\omega_{l}^{2}}{1+\tau_{p}\omega_{l}}+\gamma\omega_{l}\right)}\nonumber\\
&-4V_0\xi\left(\frac{\xi}{\beta}-1\right)\;.\label{Eq:action2}
\end{align}
In the limit of vanishing temperature  the sum over Matsubara frequencies translates into an integral $\frac{1}{\beta}\sum_{l=1}^{\infty}\rightarrow\frac{1}{2\pi}\int_{0}^{\infty}d\omega $. After a few steps of calculation we get 
\begin{align}
\frac{S_{cl,2}}{\hbar}&=\frac{\epsilon_0}{\hbar\omega_{0}}+\mathcal{I}(\gamma,\tau_p,\xi)\;,\label{Eq.Classicalaction}
\end{align}
where $S_{cl,2}=S[x_{cl}^{(2)}]$.  The first part does not depend on the distance $\xi$ and reads
\begin{equation}
\frac{\epsilon_0}{\hbar \omega_0}=\frac{8}{\pi }\frac{V_0}{\hbar\omega_0}\left[\frac{\gamma}{\omega_{0}}\left(C-\frac{\ln(\sigma)}{2}\right)-s\ln\left(\frac{\Lambda_{1}}{\Lambda_{2}}\right)\right]\;, \label{Eq.fullchemical}
\end{equation}
 with the Euler constant $C$ and $\sigma=1+\gamma\tau_p$. 
 We also defined 
 \begin{align}
s=\frac{\left(\frac{\gamma}{\omega_{0}}\Gamma_{-}-1\right)}{2\sqrt{\Gamma_{-}^{2}-1}} 
\end{align} 
and 
\begin{align}
\Lambda_{1,2} =&\frac{\omega_{0}}{\sigma}\left(\Gamma_{+}\pm\sqrt{\Gamma_{-}^{2}-1}\right)\;, \label{Eq.roots}
\end{align}
with $\Gamma_{\pm}=(\gamma\pm\tau_{p}\omega_{0}^{2})/2\omega_{0}$.
The second part of Eq.~(\ref{Eq.Classicalaction}) containing $\mathcal{I}(\gamma,\tau_p,\xi)$ depends on the distance between the two instantons $\xi$. We discuss the most general case of this part in App.~\ref{App.dilute}. For a long bounce with $\xi\omega_0\gg1$ we find 
\begin{align}
\mathcal{I}(\gamma,\xi)&\approx\frac{\eta}{\omega_0}\ln(\omega_{0}\xi) +O(1/(\xi\omega_0)^2)
\label{Eq.largebounce2}
\end{align}
where
\begin{align}
\eta=\frac{8}{\pi}\frac{V_{0}}{\hbar\omega_{0}}\gamma\;.\label{Eq.Interaction}
\end{align} 
The action on the classical path is generally of the same form as in the pure conventional dissipative case, but has a renormalized quantity $\epsilon_0$ due to momentum dissipation. We show examples of $S_{cl,2}$ in Fig.~\ref{Fig.interaction} in App.~\ref{App.dilute}.

\subsubsection{The prefactor for one instanton bounce}
%To perform 
For the calculation of the prefactor we follow the procedure given in \cite{Grabert:1987}.  Thus, we generalize the treatment of Weiss et al. to the case of two dissipative baths.
For the integration of the cyclic path integral in Eq.~(\ref{Eq.prefactornodis}), we diagonalize the action $S_\delta$  in Eq.~(\ref{Eq.deltaaction}) containing the second derivative of the time dependent potential in Eq.~(\ref{Eq.potential})
\begin{equation}
\left.\frac{d^{2}V[x(\tau)]}{dx^{2}}\right|_{x^{(2)}_{cl}}=\mathcal{V}(\tau)=\frac{m\omega_{0}^{2}}{2}\left(2-4\delta(x^{(2)}_{cl}(\tau))a_{0}\right).
\end{equation}
%The problem is now of the type of a Schr\"odinger equation describing a particle in a delta potential. 
We do this via the ansatz $ \delta x(\tau) = \sum_{q=0}^\infty c_q y_q(\tau)$, where $ y_q(\tau) $ are the eigenfunctions of the secular equation
\begin{align}
\lambda^{(2)}_{q}y_{q}(\tau)&=\left(-m\frac{d^{2}}{d\tau^{2}}+\mathcal{V}(\tau)\right)y_{q}(\tau)\nonumber\\&
+\int_{-\frac{\beta}{2}}^{\frac{\beta}{2}}d\tau'F(\tau-\tau')y_{q}(\tau')\nonumber\\&
-\int_{-\frac{\beta}{2}}^{\frac{\beta}{2}}d\tau'\widetilde{F}(\tau-\tau')\frac{d^{2}}{d\tau'^{2}}y_{q}(\tau') \label{Eq.diff}\;,
\end{align}
yielding
\begin{align}
S_{\delta,2}=\frac{1}{2}\sum_{q=0}^\infty \lambda^{(2)}_q c_q^2\;.
\end{align}
The first line in the differential Eq.~(\ref{Eq.diff}) is  equivalent to a time dependent Schr\"odinger equation describing a particle in a delta potential. For the treatment of the full model with the dissipative parts we refer to App.~\ref{App:R}. For the ansatz for $\delta x(\tau)$ we also have to perform the substitution $\delta x(\tau) \rightarrow c_p$ in the path integral yielding
\begin{align}
\oint_{\substack{\delta x(\frac{\beta}{2})=0\\ \delta x{(-\frac{\beta}{2})}=0}}D[\delta x(\tau)]\rightarrow\prod_{q=0}^{\infty} \int_{-\infty}^{\infty}\frac{dc_q}{\sqrt{2\pi \hbar}}  \mathcal{N}\;,
\end{align}
where  $\mathcal{N}$ is the Jacobian factor for the transformation.
Due to translational invariance of the bounce position on the imaginary time axis $dx^{(2)}_{cl}/d\tau=y_0$ solves the right side of Eq.~(\ref{Eq.diff}) with eigenvalue zero  $\lambda^{(2)}_0=0$. Assuming a  bounce path in which the two instantons are very far from each other, we get a second translational invariance for $\xi$ yielding a second zero eigenvalue $\lambda^{(2)}_1=0$ with eigenfunction $dx^{(2)}_{cl}/d\xi=y_1$. The presence of the position dissipation introduces a logarithmic term in the distance between the instantons. The action now weakly depends on  $\xi$. In the limit of  $\xi\rightarrow\infty$ we are back to the case of two independent instantons. However, for finite $\xi$ we treat the function $dx^{(2)}_{cl}/d\xi$ as an approximate eigenfunction, leading to $\lambda^{(2)}_{1}\approx0$.  For a more detailed analysis of this problem we refer to the App.~\ref{App.Bouncepath} and to \cite{Grabert:1987}. The eigenvalues $\lambda^{(2)}_0$ and $\lambda^{(2)}_1$ lead to divergences in the density matrix due to their invariance in $\tau$ or the weak logarithmic dependence in $\xi$. To cure this divergences we have to eliminate $\lambda^{(2)}_0$ and $\lambda^{(2)}_1$ from the eigenvalue spectra and substitute the integral 
\begin{align}
\prod_{q=0}^{\infty} \int_{-\infty}^{\infty}\frac{dc_q}{\sqrt{2\pi \hbar}}   \rightarrow  \int_{-\beta/2}^{\beta/2} \frac{d\tau_0}{A_{\tau_0}}  \int_{0}^{\beta} \frac{d\xi}{A_\xi} \prod_{q=2}^{\infty} \int_{-\infty}^{\infty}\frac{d{c}_q}{\sqrt{2\pi \hbar}}\;,
\end{align}
where $ 1/A_{\tau_0}={\left|\frac{dc_{0}(\tau_0)}{d\tau_0}\right|}$ and $ 1/A_{\xi}=\left|\frac{dc_{1}(\xi)}{d\xi}\right|$ originate from the second Jacobian transformation. We hence restored the translational invariance with the integrations over $\tau_0$ and $\xi$. 

\begin{figure}[t]
	\centering
	\includegraphics[width=1\linewidth]{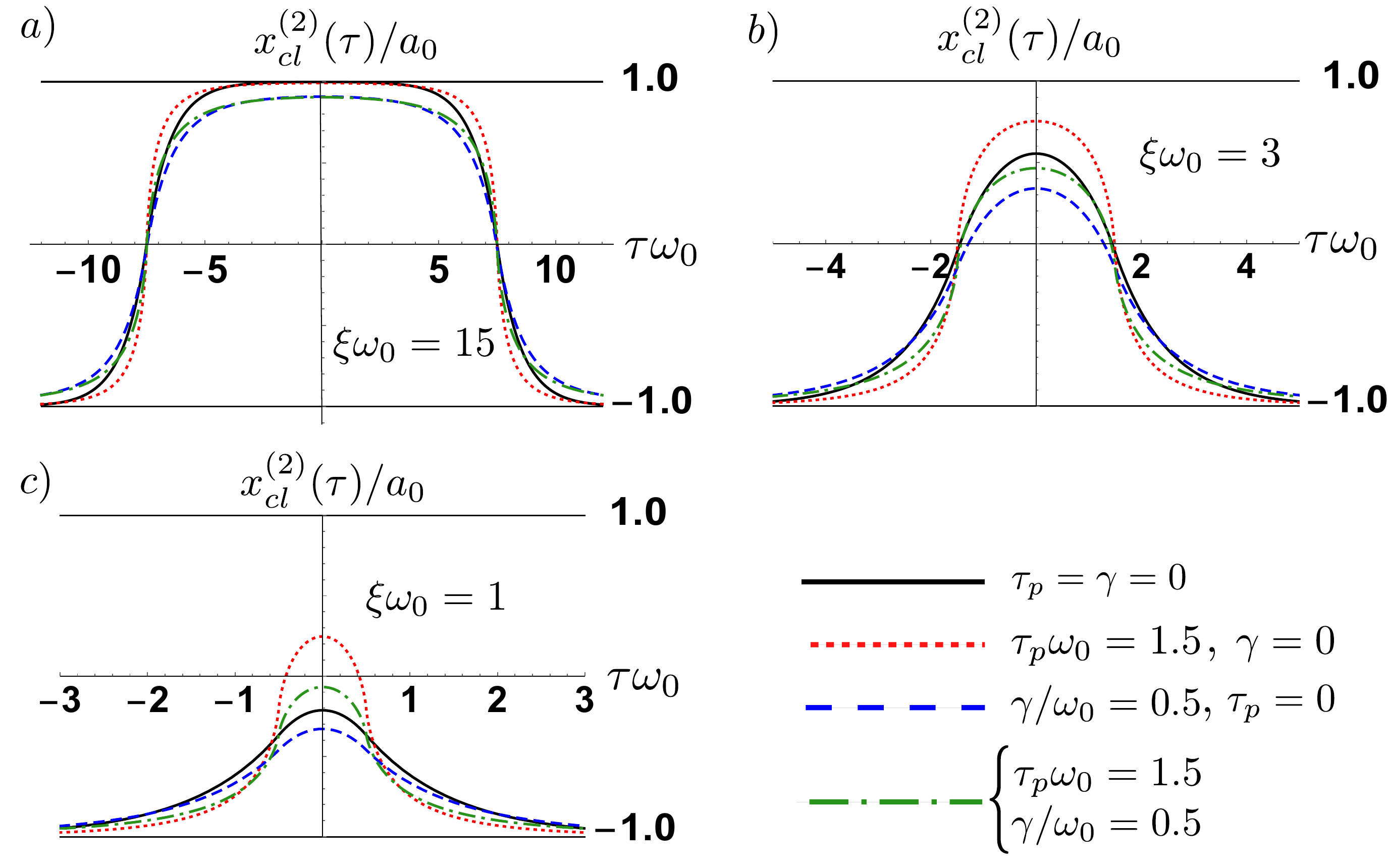}
	\caption{Path for one bounce. a) Extended bounce for different dissipative couplings. The non dissipative bounce reaches $a_0$ exponentially in $\xi\omega_0$, while the unconventional bounce reaches $a_0$ with $1/(\xi\omega_0)^2$. The conventional bounce does only reach $a_0$ with $1/(\xi\omega_0)$. b) and c) Annihilation of two instantons by reducing the spacing for different dissipative cases.}
	\label{Fig.bouncepathdiss}
\end{figure}

\subsubsection{The partition function element  for one bounce}
With this substitution we can integrate out the remaining ${c}_q$ and finally find the integral equation for the density matrix element of one bounce path
\begin{align}
z_L^{(2)}=\frac{\mathcal{N}}{2\pi\hbar}\int_{-\frac{\beta}{2}}^{\frac{\beta}{2}} \frac{d\tau_0}{A_{\tau_0}} \int_{0}^{\beta}  \frac{d\xi}{A_{\xi}}\prod_{q=2}^{\infty} \sqrt{\frac{1}{\lambda_q^{(2)}}} e^{-\frac{1}{\hbar}S_{cl,2}}\;.\label{Eq.singlepart}
\end{align}
We further rewrite Eq.~(\ref{Eq.singlepart}) by introducing the determinant for the particle staying in the left well $\prod_{q=0}^{\infty} \lambda_p^{(0)}$, and define the partition function element for zero bounces $z_L^{(0)}=\mathcal{N}/\sqrt{\prod_{q=0}^{\infty} \lambda_p^{(0)}}$, corresponding to the partition function of a particle in the harmonic potential  $V=m\omega_0^2x^2/2$ affected by dissipation. We introduce the ratio of determinants $R_B=\prod_{q=2}^{\infty} \sqrt{\lambda_q^{(0)}/\lambda_q^{(2)}}$ and the Jacobian prefactor   $\mathcal{L}_B=\frac{1}{A_{\xi}}\frac{1}{A_{\tau_0}}$ of a bounce path obtaining
\begin{align}
z_L^{(2)}=\frac{z_L^{(0)}}{2\pi\hbar} \int_{-\frac{\beta}{2}}^{\frac{\beta}{2}} d\tau_0 \int_{0}^{\beta}  d\xi\;\mathcal{L}_BR_B e^{-\frac{1}{\hbar}S_{cl,2}}\;.\label{Eq.singlepart2}
\end{align}
The quantities $R_B$ and $\mathcal{L}_B$ originate from the prefactor of one bounce. We discuss their calculation in App.~\ref{App:R} and \ref{App.L} in detail. Note that they generally depend on the distance of the bounce $\xi$. We insert the action on the classical path to obtain

%FIG.3%%%%%%%%%%%%%%%%%%%%%%%%%%%%%%%%%%%%%%%%%
\begin{figure}[t]
	\centering
	\includegraphics[width=1\linewidth]{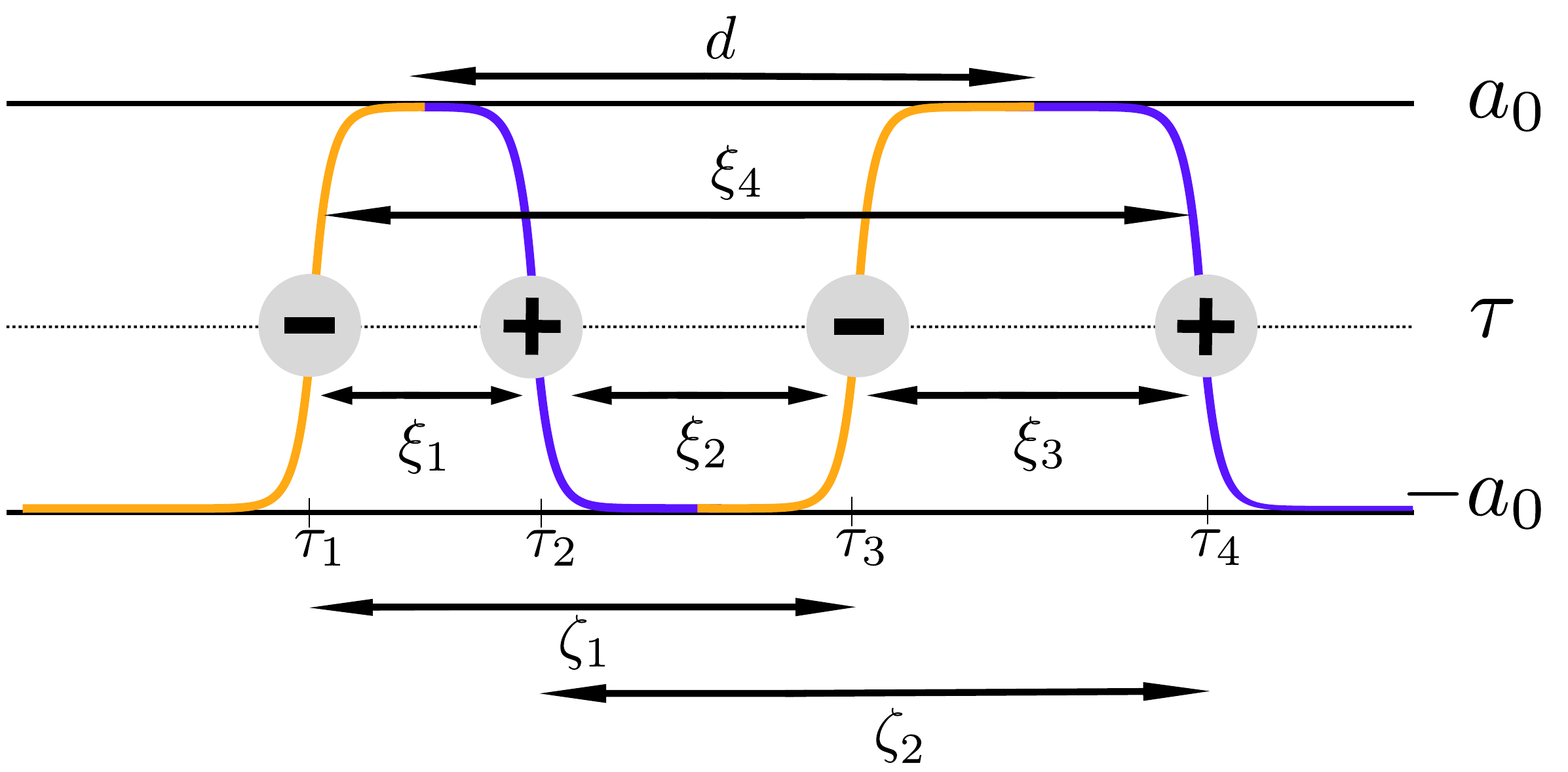}
	\caption{Path for two bounces. In presence of position dissipation the instantons behave like charges on the imaginary time axis.}
	\label{Fig.twobounces}
\end{figure}
%%%%%%%%%%%%%%%%%%%%%%%%%%%%%%%%%%%%%%%%%

\begin{align}
z_L^{(2)}=z_L^{(0)}\int_{-\frac{\beta}{2}}^{\frac{\beta}{2}} d\tau_0 \int_{0}^{\beta}  d\xi \; \mathcal{A}_B \;e^{-\mathcal{I}(\gamma,\tau_p,\xi)}\;,\label{Eq.singlepart3}
\end{align}
 where we introduced the activity 
\begin{align}
\mathcal{A}_B=\frac{R_{B}  \mathcal{L}_B e^{-\frac{\epsilon_0}{\hbar\omega_0}}}{\sqrt{2\pi\hbar}} \label{Eq.activity1}
\end{align}
of one bounce. Eq.~(\ref{Eq.singlepart3}) is the term for a bounce path starting and ending in the left well.It contains paths starting and ending at $-a_0$ meaning that also paths that do not cross the barrier (or are mostly under the barrier in the right region) are included ($\xi\omega_0 \approx 1$, see Fig.~\ref{Fig.bouncepathdiss}). Because we are interested in the delocalization of the particle we define a modified partition function element $\widetilde{z}^{(2)}$ that only contains "extended bounces" reaching deep into the other well ($\xi\omega_0\gg1$).  These bounces only depend logarithmically on their spacing to order $O(1/(\xi\omega_0)^2)$ (see App.~\ref{App.dilute}). Therefore we only use the logarithmic $\xi$ dependent part defined in Eq.~(\ref{Eq.largebounce2}) and introduce a hard core $\bar{\tau}$ for each instanton to cut off the small bounces that do not contribute to the delocalization. In App.~\ref{App.dilute} we show that in this limit the prefactors do not depend on the distance and write
\begin{align}
R_B=R_\mathcal{I}^2\;\; \text{    and    }\;\;  \mathcal{L}_B=\mathcal{L}_\mathcal{I}^2\;, \label{Eq.factorization}
\end{align}
where $R_\mathcal{I}$ and $\mathcal{L}_\mathcal{I}$ are the respective prefactors calculated for one instanton. Note that, although the action on the classical path diverges for one instanton in presence of position dissipation, we still can calculate the fluctuation prefactors ($\mathcal{L}_\mathcal{I},R_\mathcal{I}$).  Inserting Eq.~(\ref{Eq.largebounce2}) we finally obtain
\begin{align}
\widetilde{z}^{(2)}\approx  z_L^{(0)}\mathcal{A}^2_\mathcal{I} \int_{-\frac{\beta}{2}}^{\frac{\beta}{2}} d\tau_0 \int_{\bar{\tau}}^{\beta}  d\xi  \;e^{-\frac{1}{\hbar}\frac{\eta}{\omega_0}\ln(\omega_{0}\xi)}\;,\label{Eq.singlepart4}
\end{align}
where
\begin{align}
\mathcal{A}_\mathcal{I}=\frac{R_{\mathcal{I}}  \mathcal{L}_\mathcal{I} e^{-\frac{\epsilon_0}{2\hbar\omega_0}}}{\sqrt{2\pi\hbar}}\equiv e^{\frac{\mu}{\hbar \omega_0}} \label{Eq.activity2}
\end{align}
is the activity for one instanton with the quantity
\begin{align}
\mu=\left[\hbar\omega_0\ln\left(\frac{\mathcal{L}_\mathcal{I}R_\mathcal{I}}{\sqrt{2\pi \hbar}}\right)-\frac{\epsilon_0}{2}\right].
\end{align}
Now the partition function element in Eq.~(\ref{Eq.singlepart4}) is of the form of two logarithmically interacting charges on a one dimensional line with chemical potential $\mu$. The chemical potential  describes  the amount of energy an instanton/charge needs to be created. Because $\mu <0$, (i.e. it is the needed energy to create an instanton) the activity is small in the semiclassical limit and an instanton (or a bounce) is a rare event. The approximation we made above can be brought in the following physical context: We do not consider the annihilation of charges that are very close to each other. The charges we consider have a hard core of size $\bar{\tau}$. To produce one bounce/dipole out of the vacuum we have to overcome the chemical potential $2\mu$. The non trivial lowest energy contribution in the partition function is a bounce path with spacing $\bar{\tau}$. We further plot the change in the action on the classical path in App.~\ref{App.dilute}.
% Without this approximation there would be the possibility of bounces with spacing $\xi\omega_0<1$. \\ 
We  point out that, although Eq.~(\ref{Eq.singlepart2}) is of the same form as in \cite{Grabert:1987}, $S_{cl,2}$, $\mathcal{L}_\mathcal{I}$ and  $R_\mathcal{I}$ are renormalized due to the dissipative term in momentum in Eq.~(\ref{Eq.dissaction})  as we see in Sec.~\ref{Sec.results}. 

\subsubsection{The partition function element for two bounces}
To obtain $Z_L$ we sum all even $n$ instanton addends $z_L^{(n)}$. The path $x_{cl}^{(4)}$ is then the next contribution via $z_L^{(4)}$. The ansatz for the two bounces reads
\begin{align}
x_{cl}^{(4)}(\tau)&=-a_0+x_{cl}^{(1)}\left(\tau+\frac{\xi_{1}}{2}\right)-x_{cl}^{(1)}\left(\tau-\frac{\xi_{1}}{2}\right)\nonumber\\&+x_{cl}^{(1)}\left(\tau-d+\frac{\xi_{3}}{2}\right)-x_{cl}^{(1)}\left(\tau-d-\frac{\xi_{3}}{2}\right),
\end{align}
where $\xi_{1}$ is the distance between the first two instantons and $\xi_3$ is the spacing of the second bounce. Further, $d=\xi_2+\frac{\xi_1+\xi_3}{2}$ is the distance between the center of both bounces (see Fig.~\ref{Fig.twobounces}).  We insert this path into the action and find that it separates into
\begin{align}
S_{cl,4}&=S_{cl,2}[\xi_{1}]+S_{cl,2}[\xi_{2}]+S_{cl,2}[\xi_{3}]+S_{cl,2}[\xi_{4}]\nonumber\\&-S_{cl,2}[\zeta_{1}] -S_{cl,2}[\zeta_2]\;, \label{Eq.4bounceaction}
\end{align}
where $\xi_i$ denote all distances for the attractive interactions and $\zeta_i$ the ones for all repulsive interactions, respectively. Using Eq.~(\ref{Eq.Classicalaction}), we find
\begin{align}
S_{cl,4}= 2\frac{\epsilon_0}{\hbar \omega_0}+\sum_{i=1}^{4}\mathcal{I}[\xi_{i}]-\sum_{j=1}^{2}\mathcal{I}[\zeta_{j}]\;.
\end{align}
Until now we used the relative distances between the bounces. We can rewrite the interaction parts by using the imaginary time coordinates defined in Fig.~\ref{Fig.twobounces} and obtain 
\begin{align}
\sum_{i=1}^{4}\mathcal{I}[\xi_{i}]-\sum_{j=1}^{2}\mathcal{I}[\zeta_{j}]&=\mathcal{I}[|\tau_1-\tau_2|]-\mathcal{I}[|\tau_1-\tau_3|]\nonumber\\&+\mathcal{I}[|\tau_1-\tau_4|]+\mathcal{I}[|\tau_2-\tau_3|]\nonumber\\&-\mathcal{I}[|\tau_2-\tau_4|]+\mathcal{I}[|\tau_3-\tau_4|]\\&\approx
-\frac{\eta}{\omega_0}\sum_{i<j}(-1)^{j-i} \ln\left(\omega_0|\tau_i-\tau_j|\right)\;. \nonumber
\end{align}
for $\omega_0|\tau_i-\tau_j|\gg1$.
The action in Eq.~(\ref{Eq.4bounceaction}) still describes the behavior of interacting charges on a one dimensional line. To complete the discussion we deal with the prefactors for two bounces $R_{2B}$ and $\mathcal{L}_{2B}$. In the App.~\ref{App.dilute} we show that the prefactors factorize like in Eq.~(\ref{Eq.factorization}) for large spacings. Because the activity of one bounce path is very low and because bounces only weakly interact (like dipoles) it is reasonable to assume that a dilute gas of bounces with large intra-bounce distance (meaning $\xi_2\omega_0\gg 1$) forms. Hence we therefore use $R_{2B}=R_{\mathcal{I}}^4$ (and $\mathcal{L}_{2B}=\mathcal{L}_{\mathcal{I}}^4$). In this limit we obtain the partition function element for two extended bounces by distributing all four instantons on the imaginary time line, yielding
\begin{widetext}
\begin{align}
\widetilde{z}^{(4)}&\approx z^{(0)}_L\mathcal{A}_\mathcal{I}^4\; \int_{-\frac{\beta}{2}}^{\frac{\beta}{2}}d\tau_4 \int_{-\frac{\beta}{2}}^{\tau_4-\bar{\tau}}d\tau_3 \int_{-\frac{\beta}{2}}^{\tau_3-\bar{\tau}}d\tau_2 \int_{-\frac{\beta}{2}}^{\tau_2-\bar{\tau}}d\tau_1 \; e^{\frac{\eta}{\omega_0}\sum_{i<j}(-1)^{j-i} \ln\left(\omega_0|\tau_i-\tau_j|\right)}\;,
\end{align}
\end{widetext}
where we again introduced the hard core $\bar{\tau}$.

\subsubsection{Instanton gas}
Having discussed two bounce paths, the generalization  to the case of many bounces is straightforward. 
We just add  charges on the imaginary time line leading to the action on the classical path for $2k$ instantons ($k$ bounces)
\begin{align}
\frac{S_{cl,2k}}{\hbar}&=k\frac{\epsilon_0}{\hbar\omega_{0}}+\frac{\eta}{\omega_0}\sum_{i<j}^{k} (-1)^{(j-i)}\ln(\omega_{0}|\tau_i-\tau_j|)\;,
\end{align}
where the prefactor in front of the interaction guarantees the appearance of alternating charges only.  Now, dealing with an instanton gas we have to distribute all instantons on the  imaginary time line. The translational invariance of each instanton is restricted by the position of the others. Using the same arguments as before, we can write the integral equation for the diagonal density matrix element for $k$ extended bounces as
\begin{widetext}
\begin{align}
\widetilde{z}^{(2k)}_L\approx z_L^{(0)}\mathcal{A}_\mathcal{I}^{2k} \int_{-\frac{\beta}{2}}^{\frac{\beta}{2}}d\tau_{2k-1}
\int_{-\frac{\beta}{2}}^{\tau_{2k-1}-\bar{\tau}}d\tau_{2k-2} \int_{-\frac{\beta}{2}}^{\tau_{2k-2}-\bar{\tau}}d\tau_{2k-3}...\int_{-\frac{\beta}{2}}^{\tau_{1}-\bar{\tau}}d\tau_{0}\;e^{\frac{\eta}{\omega_0} \sum_{i<j}^{k} (-1)^{(j-i)}\ln(\omega_{0}|\tau_i-\tau_j|)} \label{Eq.partitionfunction}\;,
\end{align}
\end{widetext}
where we distributed all $n=2k$ instantons. 
The full  element for extended bounces $\widetilde{Z}=\sum_{k=0}^{\infty} \widetilde{z}^{(2k)}$ is now of the form of the partition function of a logarithmically interacting gas of particles with alternating charges and activity $\mathcal{A}_\mathcal{I}^{2k}$. Because $\mathcal{A}_\mathcal{I}$ is small in the semiclassical limit, 
the appearance of bounces are rare and the system forms a dilute gas of bounces. Note that momentum dissipation renormalizes the activity $\mathcal{A}_\mathcal{I}$, but does not change the general nature of the problem and the mapping to the one dimensional interacting gas is still possible. 
Anderson et al. studied such a one dimensional charge line in the context of the Kondo problem using renormalization group arguments \cite{Anderson:1970}. They predicted a phase transition for a critical parameter $\eta_c$. We use the resulting flow diagram in the next section for the study of the 
effects of the momentum dissipation on the phase transition.
%change of the critical phase transition value due to momentum dissipation. 

%RESULTS%%%%%%%%%%%%%%%%%%%%%%%%%%%%%%%%%%%%%%%%

\section{Results}
\label{Sec.results}

In this section we present our results for the unconventional dissipation, recall briefly former results for the conventional dissipation and finally discuss the frustrated case in presence of both dissipative couplings. Although we only consider extended bounces in the partition function, we start the discussion of the different dissipative cases by analyzing the modifications of the classical bounce path in presence of the respective dissipation.

\subsection{Unconventional dissipation}
We start by considering the dissipative momentum coupling only ($\gamma=0$). After shortly discussing the implications of this coupling on the bounce path trajectory, we calculate the action on the classical path and then determine the prefactor which includes the fluctuations around this path. We subsequently use this results to determine the renormalized tunneling splitting formula and discuss its implications.

\subsubsection{The classical bounce path}
We calculate the classical bounce path for unconventional dissipation by inserting $v_{l}^{(cl)}$ of Eq.~(\ref{Eq.velocityclasscialpathmatsu}) into the ansatz~(\ref{Eq.bounce}) with $\gamma=0$. Taking the zero temperature limit we find the integral
\begin{align}
x_{cl}^{(2)}=&-a_0+\frac{2\omega_{0}^{2}a_{0}}{\pi }\nonumber \\& \times \int_{0}^{\infty}\hspace{-0.2cm}d\omega \frac{\left[\sin(\omega(\tau+\frac{\xi}{2}))-\sin(\omega(\tau-\frac{\xi}{2}))\right]}{\omega\left(\omega_{0}^{2}+\frac{\omega^{2}}{1+\tau_{p}\omega}\right)}\;.
\end{align}
The numerical solution for $\tau>0$ and $\tau<0$ is shown in Fig.~\ref{Fig.bouncepathdiss}. The red dotted curve corresponds to the unconventional case. We see that the trajectory is steeper than in the non dissipative case (solid black line) meaning that the unconventional bounce can have smaller spacings before it annihilates. The bounce reaches $a_0$  asymptotically with $1/(\xi\omega_0)^2$.

\subsubsection{The action on the classical path}
For the unconventional dissipation the action on the classical path for a single extended bounce path in Eq.~(\ref{Eq.Classicalaction}) reduces to 
\begin{align}
\frac{S_{cl,2}^{(\gamma=0)}}{\hbar}=\frac{V_{0}}{\hbar\omega_{0}}\frac{4}{\pi}\frac{\ln\left(\frac{\Lambda_{1}}{\Lambda_{2}}\right)}{\sqrt{\frac{\tau_{p}^{2}\omega_{0}^{2}}{4}-1}}=\frac{\epsilon_0^{(\gamma=0)}}{\hbar \omega_0}\;, \label{Eq.unconaction}
\end{align}
where we again assumed $\xi\omega_0\gg1$. Here, the instantons do not have any long distance interaction and neglecting small bounces is equivalent to the dilute gas approximation used in the literature for the case without dissipation. It is easy to show that $S^{(\gamma=0)}_{cl,1}=\frac{1}{2}S^{(\gamma=0)}_{cl,2}$.  Fig.~\ref{Fig.Unconresults}a shows the change of the action due to the dissipative coupling $\tau_p$. The dissipation leads to a reduction of the chemical potential, which is reflected in an increased tunnel splitting as we will illustrate below.

\subsubsection{The prefactor of one bounce}
\begin{figure}[t]
	\centering
	\includegraphics[width=0.99\linewidth]{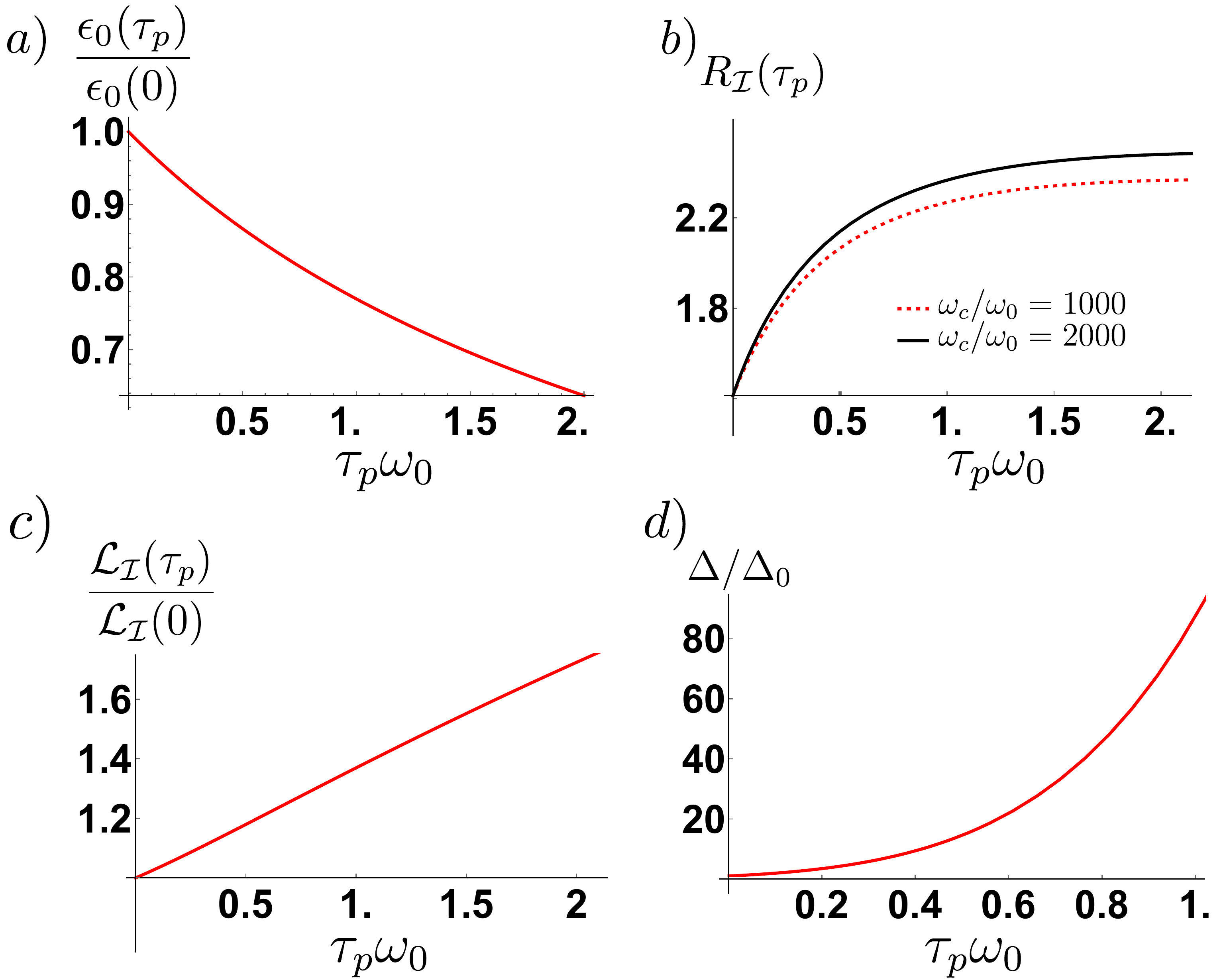}
	\caption{Results for unconventional dissipation: a) $\epsilon_0$ depending on $\tau_p$ and scaled with $\epsilon_0(0)/\hbar\omega_0=4V_0/\hbar\omega_0$. b) Ratio of the determinants for different values of $\omega_c$. c) Jacobian prefactor for one bounce as a function of the unconventional coupling $\tau_p$. d) Tunnel splitting as a function of $\tau_p$ for $V_0/\hbar\omega_0=8$, with $\Delta_0$ being the tunnel splitting in absence of dissipation.}
	\label{Fig.Unconresults}
\end{figure}

We determine the prefactor of a single bounce by calculating the ratio of the determinants $R_\mathcal{I}^{(\gamma=0)}$ and the Jacobian prefactor $\mathcal{L}^{(\gamma=0)}_\mathcal{I}$.
The detailed procedure of the calculation can be found in App.~\ref{App:R} and \ref{App.L}.
We find an ultraviolet divergence in $R_\mathcal{I}^{(\gamma=0)}$ for the unconventional ohmic dissipation, which originates from diverging position quantum fluctuations in presence of this coupling \cite{Rastelli:2016ge}.  This divergence is cured by a high frequency cutoff function $f_c=\left(1+{\omega_l}/{\omega_c}\right)^{-1}$, where $\omega_c$ is the cutoff \cite{Weiss:2012}.  
In Fig.~\ref{Fig.Unconresults}b we illustrate the change of the behavior due to the presence of unconventional dissipation. $R_\mathcal{I}^{(\gamma=0)}(\tau_p)$ increases monotonically with increasing $\tau_p$ (for the non dissipative case we find $R_\mathcal{I}^{(\gamma=0,\tau_p=0)}=\sqrt{2}$). We also show the weak cutoff dependence of this result by presenting data for different values of $\omega_c$ and proceed from now on  with $\omega_c/\omega_0=1000$. 
The remaining factor $\mathcal{L}_\mathcal{I}^{(\gamma=0)}$ contains all Jacobian factors $1/A_x$.
Fig.~\ref{Fig.Unconresults}c shows that $\mathcal{L}_\mathcal{I}^{(\gamma=0)}$ increases monotonically with $\tau_p$.

%Fig.4%%%%%%%%%%%%%%%%%%%%%%%%%%%%%%%%%%%%%%%%%%%%%%%%%%%%%%%
%%%%%%%%%%%%%%%%%%%%%%%%%%%%%%%%%%%%%%%%%%%%%%%%%%%%%%%%%

\subsubsection{The tunnel splitting with unconventional dissipation}
Inserting Eq.~(\ref{Eq.unconaction}) and the quantities $R_\mathcal{I}$ and $\mathcal{L}_\mathcal{I}$ into Eq.~(\ref{Eq.partitionfunction}),  the integrand turns out to be independent of $\tau_i$. Hence, we can perform all integrations yielding the distribution factor
\begin{align}
\int_{-\frac{\beta}{2}}^{\frac{\beta}{2}}d\tau_{2k-1}
\int_{-\frac{\beta}{2}}^{\tau_{2k-1}-\bar{\tau}}d\tau_{2k-2}...\int_{-\frac{\beta}{2}}^{\tau_{1}-\bar{\tau}}d\tau_{0}=\frac{\beta^{2k}}{2k!}\;,
\end{align}
where we used the approximation $(\beta-2k\bar{\tau})^{2k}\approx\beta^{2k}$ which is valid for small temperatures ($\beta\rightarrow \infty$). With this we get
\begin{align}
\widetilde{Z}=z_L^{(0)}\sum_{k=0}^{\infty}\left(\mathcal{A}_\mathcal{I}^{(\gamma=0)}\right)^{2k} \frac{\beta^{2k}}{2k!}\;.\label{Eq.Delta0}
\end{align}
The quantity $z^{(0)}_L$ is the semiclassical fluctuation prefactor of a harmonic oscillator with unconventional dissipation. We can rewrite it in general as
\begin{align}
z^{(0)}_L=\frac{1}{\sqrt{2\pi}}\sqrt{\frac{1}{\langle x^2 \rangle}}e^{-\frac{\beta }{2}\omega(\tau_p)}\;,
\end{align}
where $\hbar\omega(\tau_p)/2$ corresponds to the lowest energy level in each parabola and $\langle x^2 \rangle$ are the quantum position fluctuations in presence of unconventional dissipation. We then perform the summation in Eq.~(\ref{Eq.Delta0}) and finally obtain 
\begin{align}
\widetilde{Z}= \frac{1}{2\sqrt{2\pi}}\sqrt{\frac{1}{\langle x^2 \rangle}}e^{-\frac{\beta }{2}\omega(\tau_p)} \left(e^{\beta\frac{\Delta(\tau_p)}{2}}-e^{-\beta\frac{\Delta(\tau_p)}{2}}\right)\;,
\end{align}
with
\begin{align}
\frac{\Delta(\tau_p)}{2}=\mathcal{A}_\mathcal{I}^{(\gamma=0)}=\frac{R_{\mathcal{I}} (\tau_p) \mathcal{L}_\mathcal{I}(\tau_p) e^{-\frac{\epsilon_0(\tau_p)}{2\hbar\omega_0}}}{\sqrt{2\pi\hbar}}\;.
\end{align}
By comparing this expression to the spectral representation of ${Z}_L$ (reported in App.~\ref{App.Offdiag}), we find that $\Delta(\tau_p)$ is the tunnel splitting of the coherent superposition of the ground states in the left and the right well.  Fig.~\ref{Fig.Unconresults}d shows the behavior of this splitting as a function of $\tau_p$. In the limit $\tau_p=0$ we find the well-known non dissipative result for the splitting in a symmetric parabolic double well $\Delta_0=\sqrt{2\hbar\omega_0V_0/\pi}\exp({-2V_0/(\hbar\omega_0)})$, whereas the generic tunneling dependence $\Delta(\tau_p)$ exhibits an increasing behavior with the dissipation.

\begin{figure}[t]
	\centering
	\includegraphics[width=1.0\linewidth]{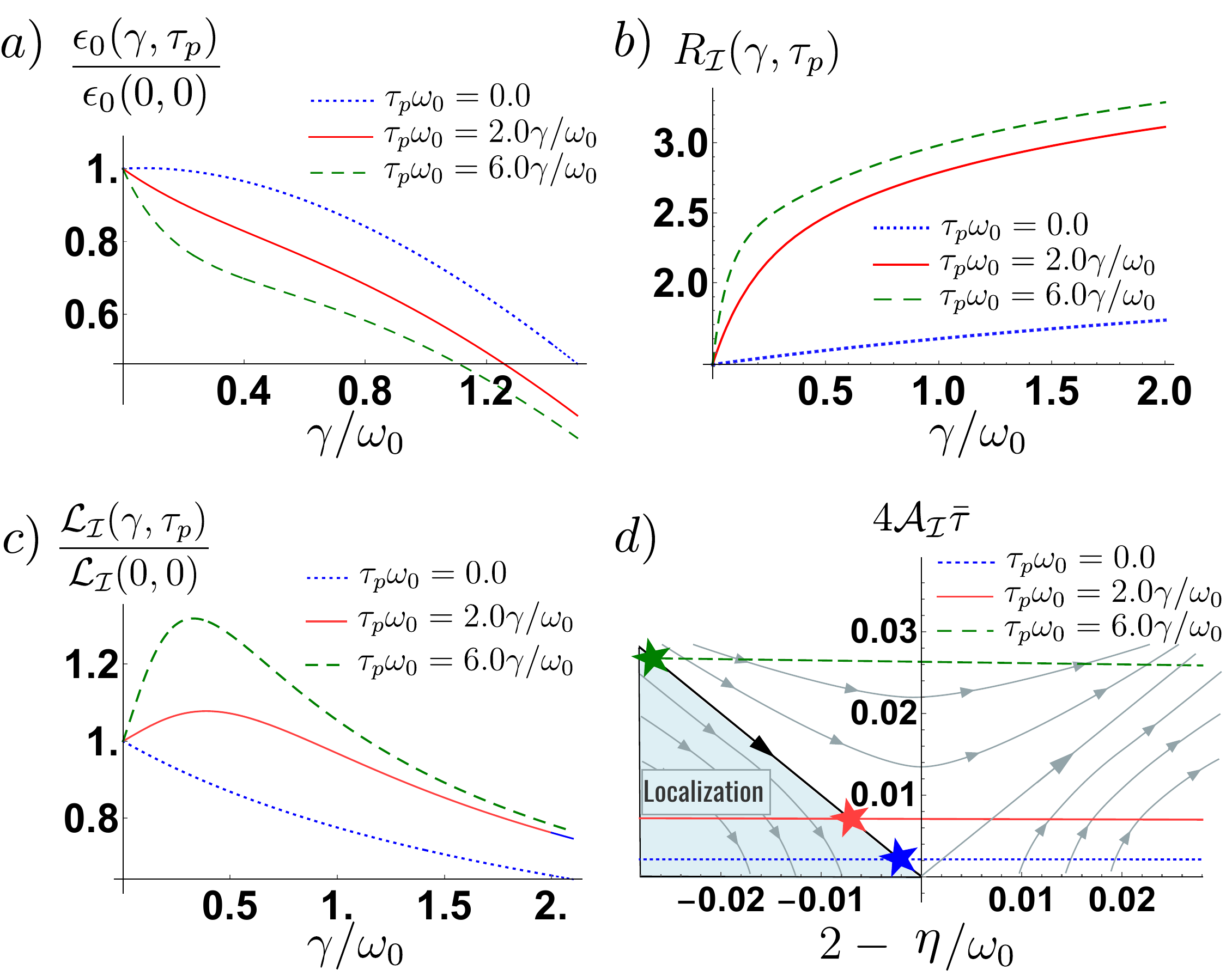}
	\caption{Results for conventional (red dotted line) and frustrated dissipation. a) Non interacting part of the action. b) Ratio of the determinants. c) Jacobian prefactor for one bounce as a function of conventional dissipation for different values of unconventional couplings. d) Flow diagram for the gas of interacting instantons. The blue shaded region in the parameter space corresponds to the localization. By changing $\gamma$ the interaction $\eta$ ($x$ axis) and the activity $\mathcal{A}_\mathcal{I}$ ($y$ axis) change. The resulting lines are plotted for $V_0/\hbar\omega_0=4$. Different lines correspond to different unconventional couplings. The stars indicate the phase transition points.}
	\label{Fig:phasechange2}
\end{figure}

\subsection{Conventional dissipation}
The effect of conventional dissipation ($\gamma\neq 0, \tau_p=0$) on quantum systems is by far more explored then the effect of the above discussed unconventional coupling. Although Caldeira and Leggett mention the possibility of momentum couplings in their early works, the quantum systems investigated theoretically are mainly coupled in the conventional sense.
Chakravarty et al. and Bray et al. studied the quartic double well  in presence of position dissipation analytically \cite{Chakravarty:1982,Bray:1982}. They mapped the problem into a gas of interacting particles and used Andersons renormalization procedure  to predict a transition from tunneling to localization by increasing the dissipative coupling strength.  Matsuo et al. determined the localization transition in a quartic double well via a numerical path integral Monte Carlo treatment \cite{Matsuo1:2004,Matsuo2:2008}. 
%%%%
The full treatment of calculating the action on the classical path and the prefactor for a bounce path in a parabolic double well is studied in \cite{Grabert:1987}. We recover the same result for $\tau_p=0$ and a single bounce. 
%%%

\subsubsection{The classical bounce path for conventional dissipation}
We can calculate the classical bounce path in the same way as in the unconventional case, with $v_{l}^{(cl)}(\tau_p=0)$. The result is shown as blue dashed line in Fig.~\ref{Fig.bouncepathdiss}. We see that the path affected by conventional dissipation only asymptotically reaches $a_0$ with $1/\xi\omega_0$. The bounce path for $\xi\omega_0 = 15$ is nevertheless an \textit{extended} bounce because the interaction between the instantons is logarithmic to order $O(1/(\xi\omega_0)^2)$ (see Fig.~\ref{Fig.interaction}a in the Appendix). By bringing the two instantons closer we see that they start to annihilate already for larger spacings $\xi$ than the non dissipative case (see Fig.~\ref{Fig.bouncepathdiss}b-c).

\subsubsection{Partition function element and phase transition}
Here we deal with the full interacting gas  $\widetilde{Z}=\sum_{k=0}^{\infty}\widetilde{z}^{(2k)}$, with $\widetilde{z}
^{(2k)}$  defined in Eq.~(\ref{Eq.partitionfunction}). We calculate the ratio of determinants $R^{(\tau_p=0)}_\mathcal{I}$ and the Jacobian prefactor $\mathcal{L}^{(\tau_p=0)}_\mathcal{I}$ in the same way as in the unconventional case. In Fig.~\ref{Fig:phasechange2}b and c we show the results for these quantities together with
\begin{align}
\frac{\epsilon_0^{(\tau_p=0)}}{\hbar\omega_0}=\frac{\eta}{\omega_{0}}C+\frac{V_{0}}{\hbar\omega_{0}}\frac{4}{\pi}\frac{\left(1-\frac{\gamma^{2}}{2\omega_{0}^{2}}\right)}{\sqrt{\frac{\gamma^{2}}{4\omega_{0}^{2}}-1}}\ln\left(\frac{\Lambda_{1}}{\Lambda_{2}}\right)
\end{align}
in Fig.~\ref{Fig:phasechange2}a (blue dotted lines).
As discussed above, the problem is of the form of a one dimensional gas of logarithmically interacting charges with the activity $\mathcal{A}_{\mathcal{I}}^{(\tau_p=0)}$. We therefore use the procedure introduced by Anderson and Yuval in \cite{Anderson:1970}. By integrating out small bounce contributions on this one dimensional charge line one finds a renormalization of the coupling $\eta$ and of the activity  $\mathcal{A}_\mathcal{I}^{(\tau_p=0)}$. The resulting flow diagram is shown in Fig.~\ref{Fig:phasechange2}d, where the points left of the line 
\begin{align}
2-\eta/\omega_0=-4\mathcal{A}_\mathcal{I}^{(\tau_p=0)}\bar{\tau}
\end{align}
in parameter space (shaded region) scale  to zero activity, meaning infinite chemical potential $\mu=-\infty$. In this region no bounce appears and the particle is localized in the left well. The critical line depends on the small distance cutoff $\bar{\tau}$. For simplicity we choose $\bar{\tau}\approx1/\omega_0$ from now on. So the closest configurations we account for are purely logarithmically interacting instantons with spacings $|\tau_i-\tau_j|\omega_0\approx1$. 
In absence of the latter, all other non trivial configurations do not appear. In the limit $V_0/\hbar\omega_0\rightarrow \infty$, we infer from Fig.~\ref{Fig:phasechange2}d that $\eta_c/\omega_0=2$, which implies that due to Eq.~(\ref{Eq.Interaction}) an infinitesimal small dissipative coupling $\gamma$ already leads to localization in this limit \footnote{We use $V_0/\hbar\omega_0=4$ in Fig.~\ref{Fig:phasechange2}d to illustrate the change of the activity due to the unconventional coupling. The $V_0$ dependence of the result is plotted in Fig.~\ref{Fig:phasechange}a-b. }.
%For a very deep potential $V_0\rightarrow \infty$ the activity is almost zero and we can read of from Fig.~\ref{Fig:phasechange2}d that $\eta_c=2$ and therefore $\gamma_c\rightarrow 0$.
 %
In Fig.~\ref{Fig:phasechange}b we show the phase diagram for the interaction $\eta_c$ as a function of $V_0$ (black lines). The phase diagram for the interaction can be compared to one of Matuso et al. \cite{Matsuo1:2004} in the regime $V_0/\hbar\omega_0\gg1$.
The change of the critical value $\eta_c$ by decreasing $V_0$ is purely due to the increasing activity (note that the $V_0$ dependence of $\eta$ does not enter here).
The phase diagram for the dissipative coupling $\gamma_c$ as a function of $V_0$ is reported in Fig.~\ref{Fig:phasechange}a. The black solid line corresponds to the conventional dissipative case. The particle is delocalized below this line, above the particle is trapped in the well.

\subsection{Conventional and unconventional dissipation}
\subsubsection{The classical bounce path with both dissipative couplings}
The classical bounce path for this case is also shown in Fig.~\ref{Fig.bouncepathdiss}. For an extended bounce it almost coincides with the one of purely conventional dissipation, while for small bounces the effect of unconventional dissipation becomes important. Here the path leaks further into the right well (and annihilates later) as in the conventional case for $\xi\omega_0\approx3$. For $\xi\omega_0\approx 1$ the bounce is even more extended than the non dissipative one.

\subsubsection{Change in the phase diagram due to the unconventional dissipation}

\begin{figure}[t]
	\centering
	\includegraphics[width=1.0\linewidth]{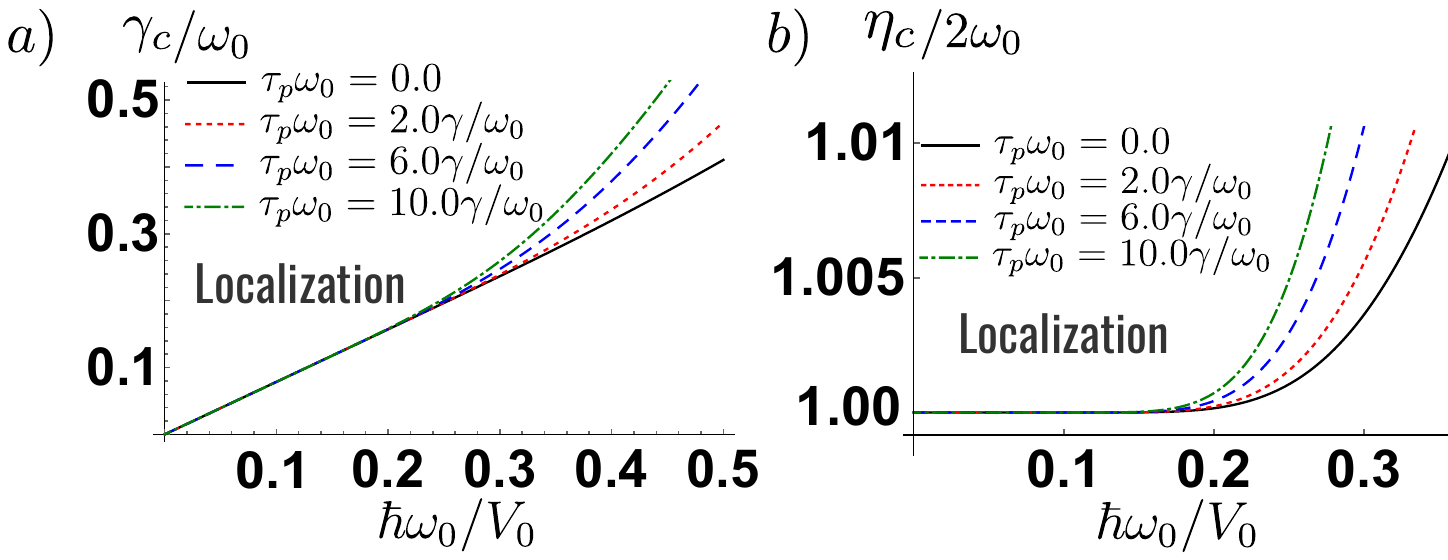}
	\caption{Critical value for the conventional coupling as a function of $V_0$.  Different lines correspond to different ratios $\tau_p \omega_0^2/\gamma$. a) Critical value for $\gamma$. b) Critical value for $\eta$.}
	\label{Fig:phasechange}
\end{figure}

We finally discuss how the enhancement of the splitting due to unconventional dissipation affects the phase transition determined by the conventional dissipation. From Eq.~(\ref{Eq.Classicalaction}) we see that the unconventional part does not affect the long-range interaction between the instantons for extended bounces. However, the activity $\mathcal{A}_\mathcal{I}(\gamma,\tau_p)$ is affected by it, as illustrated in Fig.~\ref{Fig:phasechange2}a - d. In particular, the decrease of $\epsilon_0$ is more pronounced in presence of unconventional dissipation. For a sufficiently large momentum dissipation, $\mathcal{L}_\mathcal{I}$ exhibits a non monotonic behavior as a function of the dissipative couplings (see Fig.~\ref{Fig:phasechange2}c), while $R_\mathcal{I}$ increases more for large values of $\tau_p$, see Fig.~\ref{Fig:phasechange2}b. The phase transition is determined by the condition $2-\eta/\omega_0=-4\mathcal{A}_\mathcal{I}\bar{\tau}$, as shown in Fig.~\ref{Fig:phasechange2}d. The critical point in parameter space is highlighted by the stars. We find that the activity increases with $\tau_p$, leading to a higher value for the critical interaction $\eta_c$. The change in the phase transition point varies strongly with $V_0$: For a very high potential $V_0\rightarrow \infty$, the activity is suppressed by the large value of $\epsilon_0$ and $\mathcal{A}_\mathcal{I}\rightarrow0$. This means that the effect of unconventional dissipation is very small and the critical value for the interaction of the bounces $\eta_c/\omega_0=2$ independently of $\tau_p$. However, by decreasing $V_0$  the influence of the unconventional dissipation becomes stronger and the change in $\eta_c$ and therefore in $\gamma_c$ is more important. In Fig.~\ref{Fig:phasechange}b we show the phase diagram for $\eta_c$.  We find that the unconventional dissipation shifts the critical coupling to higher values. 
In Fig.~\ref{Fig:phasechange}a we report the critical values for the dissipative coupling $\gamma_c$ for different dissipative cases.  
%We see that for large values of  $V_0$ the critical coupling $\gamma_c$ is very small.
% It increases by decreasing $V_0 $. 
Reducing $V_0$, the unconventional dissipation becomes more and more important, and increasing the ratio between both dissipative couplings leads to larger critical values and consequently to a more extended delocalized phase. 
By decreasing the ratio $V_0/\hbar\omega_0$, we approach the regime in which the semiclassical approximation does not hold a priori. The behavior of the critical line in Fig.~\ref{Fig:phasechange} suggests however, that in this regime the unconventional dissipation may have a stronger impact on the quantum phase transition as compared to the weak renormalization observed in the present study. 

%FIG.6%%%%%%%%%%%%%%%%%%%%%%%%%%%%%%%%%%%%%%%%%%%%%%%
%%%%%%%%%%%%%%%%%%%%%%%%%%%%%%%%%%%%%%%%%%%%%%%%%%%%

%CONCLUSION%%%%%%%%%%%%%%%%%%%%%%%%%%%%%%%%%%%%%%%%%%%%
\section{Conclusions}
\label{Sec.Conclusion}
We studied a dissipative quantum system formed by a particle moving in a double well and coupled to two different baths via non commuting operators ($\hat{p}$ and $\hat{x}$).
We use the Euclidean path integral method to analyze the problem in the semiclassical limit at zero temperature.
First we investigate the effect of a single dissipative momentum coupling on a particle in a double well potential.  We find that this coupling enhances the tunnel splitting favoring the quantum behavior of the system.
In contrast, position dissipation leads to a localization phase transition by increasing the dissipative coupling. Hence, the two  dissipative couplings can exhibit competing effects. 
We find that the momentum coupling  affects the localization transition originating from conventional position 
dissipation. The momentum-induced reduction of the chemical potential leads to a shift of the critical conventional coupling to higher values. 
%The influence of unconventional dissipation depends on the activity $\mathcal{A}_\mathcal{I}$ and therefore on the height $V_0$ of the potential barrier. 
This can be related to the potential barrier height: 
In the limit of a very high barrier the particle is strongly confined and the influence of the momentum coupling is very small. However, for decreasing $V_0$ we find that momentum coupling increases the critical value $\gamma_c$ substantially, 
due to the increasing importance of quantum fluctuations. We argue that in the region $\hbar\omega_0/V_0\gtrsim1$ not accessibile within the present semiclassical description, momentum dissipation  can lead to an even more substantial change of the phase diagram. 
Finally, dissipative momentum couplings can be translated into charge couplings in electrical quantum systems. Therefore, superconducting junctions are  possible platforms to test  the effect of unconventional dissipative couplings \cite{Maile:2018}.

%%%%%%%%%%%%%%%%%%%%%%%%%%%%%%%%%%%%%%%%%%%%%%%

\acknowledgments
The authors thank Wolfgang Belzig for valuable discussions. We acknowledge
financial support from the Deutsche Forschungsgemeinschaft (DFG) through ZUK 63 and the Zukunftskolleg, and by the MWK-RiSC program.

\appendix

\section{Off-diagonal density matrix elements}
\label{App.Offdiag}
We start by considering two independent parabolas with ground states $|L_0\rangle$ and $|R_0\rangle$. Forming the double well potential Fig.~\ref{Fig.Instanton}a with this two parabolas a particle that is localized in one of the wells starts to tunnel through the barrier $V_0$. Due to this tunneling the particle interferes with itself leading  to a superposition of the two states $|L_0\rangle$ and $|R_0\rangle$ and to the new ground states of the combined system
\begin{equation}
|\psi_{\pm}\rangle \cong \frac{1}{\sqrt{2}}\left(|L_0\rangle \pm |R_0\rangle\right)
\end{equation}
with respective energies $E_{\pm}=\hbar \omega_0 \mp \frac{\hbar\Delta}{2}$, where $\Delta$ is the tunnel splitting. 
The density matrix element $\rho_{-a_0,-a_0}$ in the spectral representation then reads
\begin{align}
\rho_{-a_{0},-a_{0}}&=\langle-a_{0}|e^{-\frac{\beta\hat{H}}{\hbar}}|-a_{0}\rangle=\sum_{n}e^{-\frac{\beta E_{n}}{\hbar}}\langle-a_{0}|n\rangle\langle n|-a_{0}\rangle\nonumber\\ &=e^{-\frac{\beta E_{+}}{\hbar}}\langle-a_{0}|\psi_{+}\rangle\langle\psi_{+}|-a_{0}\rangle\nonumber\\&-e^{-\frac{\beta E_{-}}{\hbar}}\langle-a_{0}|\psi_{-}\rangle\langle\psi_{-}|-a_{0}\rangle\;,
\end{align}
where we considered only the two lowest eigenvalues since we are in the semiclassical limit and at low temperature.

\section{Bounce path calculation}
\label{App.Bouncepath}
\subsection{Variable transformation}
We follow in the discussion of the prefactor $R_B$ Refs.~\cite{Grabert:1987,Kleinert:2009}
and start by considering the path $x^{(2)}(\tau,\xi) $, i.e. one bounce as shown in Fig.~\ref{Fig.Instanton}c. The position of the bounce is denoted by $\tau_0$ and the spacing between the instantons by $\xi$. For the solutions of the differential Eq.~(\ref{Eq.diff}) there exists a zero eigenvalue $\lambda^{(2)}_0=0$ for the eigenfunction $y_0(\tau)=A_{\tau_0}\dot{x}_{cl}(\tau)$ because of translational invariance of the bounce on the whole $\tau$ axis.  $A_{\tau_0}$ is constant and $\dot{x}_{cl}(\tau)=d{x}_{cl}(\tau)/d\tau$. Further, the spectrum contains a small negative eigenvalue $\lambda^{(2)}_1\lesssim0$ due to  the dissipation induced interaction between the instantons forming the bounce. Therefore the parameter $\xi$ is the relevant quantity in this case. These two eigenvalues lead to divergences in the partition function element. We get rid of the two eigenvalues $\lambda^{(2)}_0$ and $\lambda^{(2)}_1$ via changes of variables $c_0 \rightarrow \tau_0$ and $c_1 \rightarrow \xi$ with the ansatz
\begin{equation}
x^{(2)}(\tau,\xi )=x^{(2)}_{cl}(\tau-\tau_0,\xi)+\sum_{q=2}^{\infty } c_q y_q(\tau-\tau_0,\xi)\;.\label{Eq.Fullpath}
\end{equation}
As a consequence of translational invariance, we have for the action on the classical path $ S_{cl,2}(x_{cl,2}(\tau-\tau_0,\xi))=S_{cl,2}(x_{cl}^{(2)}(\tau,\xi))=S_{cl,2}(\xi)$. The Jacobian $|dc_0/d\tau_0|$  of the variable transformations is obtained by the overlap of $x^{(2)}(\tau,\xi )$ with $y_0(\tau)=A_{\tau_0}\dot{x}_{cl}(\tau)$ yielding
\begin{align}
c_0(\tau_0)&=\int_{-\frac{\beta}{2}}^{\frac{\beta}{2}}x^{(2)}(\tau,\xi )y_0(\tau,\xi)\nonumber \\
&=\int_{-\frac{\beta}{2}}^{\frac{\beta}{2}}d\tau[x_{cl}^{(2)}(\tau-\tau_{0},\xi)A_{\tau_{0}}\dot{x}_{cl}^{(2)}(\tau,\xi)\nonumber\\&+\sum_{q=2}^{\infty}c_{q}y_{q}(\tau-\tau_{0},\xi)y_{0}(\tau)]\;.\label{Eq.co1}
\end{align}
Because of translational invariance we can expand %the functions
\begin{align}
x_{cl}^{(2)}(\tau-\tau_{0},\xi)&\approx x_{cl}^{(2)}(\tau,\xi)+\dot{x}_{cl}^{(2)}(\tau,\xi)\tau_{0}
\end{align}
and analogously $y(\tau-\tau_{0},\xi) \approx y(\tau,\xi)+\dot{y}(\tau,\xi)\tau_{0}$. Inserting these expressions into (\ref{Eq.co1}) 
we find $(x_{cl}^{(2)}=x_{cl,2})$
\begin{align}
c_0(\tau_0)&= \int_{-\frac{\beta}{2}}^{\frac{\beta}{2}}d\tau[A_{\tau_{0}}\left(x_{cl,2}(\tau,\xi)\dot{x}_{cl,2}(\tau,\xi)+\dot{x}_{cl,2}^{2}(\tau,\xi)\right)\nonumber\\&+\sum_{q=2}^{\infty}c_{q}\dot{y}_{q}(\tau,\xi)y_{0}(\tau)]\tau_{0}
\end{align}
and hence
\begin{align}
\left|\frac{dc_{0}(\tau_{0})}{d\tau_{0}}\right|
&=\frac{1}{A_{\tau_{0}}}+A_{\tau_{0}}\int_{-\frac{\beta}{2}}^{\frac{\beta}{2}}d\tau\sum_{q=2}^{\infty}c_{q}\dot{y}_{q}(\tau,\xi)\dot{x}_{cl,2}(\tau,\xi)\;,\label{Eq.Jac1}
\end{align}
where we used that 
\begin{equation}
A_{\tau_{0}}=\left(\sqrt{\int_{-\beta/2}^{{\beta/2}}d\tau\dot{x}_{cl,2}^{2}(\tau)}\right)^{-1}\;,
\end{equation}
which follows from the normalization condition $\int_{-\beta/2}^{{\beta/2}}y^2_0(\tau)d\tau=1$. The small negative eigenvalue $\lambda^{(2)}_1$ can be dealt with in a similar manner. \\
The classical path depends logarithmically on the spacing $\xi$ between instantons. So the system only weakly depends on changes of $\xi$, which makes 
\begin{equation}
y_1(\tau,\xi)=A_\xi \frac{d}{d\xi}x_{cl,2}(\tau,\tau_0,\xi)=A_\xi x'_{cl,2}(\tau,\tau_0,\xi)
\end{equation}
an "approximate" eigenfunction of the differential Eq. (\ref{Eq.diff}). Treating the problem in the same way as the zero mode we deal with a spacing of the bounce $\xi\omega_0\gg1$ and introduce the small change  $d\xi$ to obtain 
\begin{align}
x_{cl,2}(\tau,\tau_0,\xi-d\xi)&\approx x_{cl,2}(\tau,\tau_0,\xi)+x'_{cl,2}(\tau,\tau_0,\xi)d\xi
\end{align}
and $y(\tau,\tau_0,\xi-d\xi) \approx y(\tau,\tau_0,\xi)+y'(\tau,\tau_0,\xi)d\xi$. We hence find for the coefficient
\begin{align}
dc_1(\xi)&= \int_{-\frac{\beta}{2}}^{\frac{\beta}{2}}d\tau[A_\xi x'^2_{cl,2}(\tau,\tau_0,\xi)\nonumber\\&+A_\xi\sum_{q=2}^{\infty}c_{q}y'_q(\tau,\tau_0,\xi)x'_{cl,2}(\tau,\tau_0,\xi)]d\xi \;.
\end{align}
Using the normalization condition $\int_{-\beta/2}^{{\beta/2}}y^2_1(\tau)d\tau=1$ to determine 
\begin{equation}
A_{\xi}=\left({\sqrt{\int_{-\beta/2}^{{\beta/2}}d\tau x_{cl,2}^{'2}(\tau)}}\right)^{-1} \;,
\end{equation}
yields
\begin{align}
\left|\frac{dc_{1}(\xi)}{d\xi_0}\right|
&=\frac{1}{A_{\xi}}\nonumber\\&+A_{\xi}\int_{-\frac{\beta}{2}}^{\frac{\beta}{2}}d\tau\sum_{q=2}^{\infty}c_{q}{y'}_{q}(\tau,\tau_0,\xi){x'}_{cl,1}(\tau,\tau_0,\xi)\;.\label{Eq.Jac2}
\end{align}
We therefore obtained the Jacobian transformation for the substitutions $c_0 \rightarrow \tau_0$ and $c_1 \rightarrow \xi$. 

\subsection{Transformation of the integral measure and partition function for one bounce}
We also have to deal with the integration measure of the path integral, we find for the general time sliced measure
\begin{align}
\oint_{-a_0}\mathcal{D}[x_2(\tau)]\rightarrow \lim_{M\rightarrow\infty} \prod_{i=0}^{M-1 } \int_{-\infty}^{\infty}\frac{dx_{2,i}}{\sqrt{2\pi \hbar}}\;,
\end{align}
where $x_0=x_{M-1}=-a_0$. We use the (diagonal) ansatz $x_{2,i} =x_{cl,2,i}+\sum_{q=0}^{\infty} c_q y_{q,i}$ for the path. The first step is to transform to
\begin{align}
\lim_{M\rightarrow\infty}\prod_{i=0}^{\infty } \int_{-\infty}^{\infty}\frac{dx_{2,i}}{\sqrt{2\pi \hbar}}=\mathcal{N} \prod_{q=0}^{\infty} \int_{-\infty}^{\infty}\frac{dc_q}{\sqrt{2\pi \hbar}}\;,\label{Eq.measure}
\end{align}
where $\mathcal{N}$ is a constant determined by $dx_{2,i}=\mathcal{N} dc_q$. However, in order to avoid the divergences originating from the eigenvalues $\lambda^{(2)}_0$ and $\lambda^{(2)}_1$, we use the path of Eq.~(\ref{Eq.Fullpath}). This implies the substitutions $c_0\rightarrow \tau_0$ and $c_1 \rightarrow \xi$, together with the according modification of the measure in Eq.~(\ref{Eq.measure}).  Further, since the action on the classical path $S_{cl,2}(\xi)$ depends on the parameter $\xi$, we cannot extract it from the integral. We choose the functions $y_q(\tau,\tau_0,\xi)$ (now $q>1$) to be eigenfunctions of the differential equation (\ref{Eq.diff}),  therefore the action on the fluctuations excluding $\lambda^{(2)}_0$ and $\lambda^{(2)}_1$  reads $\widetilde
{S}_{\delta,2} =\frac{1}{2}\sum\limits_{q=2}^{\infty} \lambda^{(2)}_q c^2_q$. With this we write for the partition function of one bounce path
\begin{widetext}
\begin{align}
z_{L,2}
&=\mathcal{N} \prod_{q=2}^{\infty} \int_{-\infty}^{\infty}\frac{dc_q}{\sqrt{2\pi \hbar}} \int_{-\beta/2}^{\beta/2} \frac{d\tau_0}{\sqrt{2\pi\hbar}} \left|\frac{dc_{0}(\tau_0)}{d\tau_0}\right| \int_{0}^{\beta} \frac{d\xi}{\sqrt{2\pi\hbar}} \left|\frac{dc_{1}(\xi)}{d\xi}\right| e^{-\frac{1}{\hbar}\frac{1}{2}\sum\limits_{q=2}^{\infty} \lambda^{(2)}_q c^2_q}\;,
\end{align}
\end{widetext}
where the product only affects the integral. The Jacobians $\left|\frac{dc_{0}(\tau_0)}{d\tau_0}\right|$ and $\left|\frac{dc_{1}(\xi)}{d\xi}\right|$ are provided in Eqs. (\ref{Eq.Jac1}) and (\ref{Eq.Jac2}) respectively. The divergence originating from the translational invariance is now removed because of the integration boundaries of the integral over $\tau_0$. We integrate out the $c_q$ and find
\begin{align}
z_{L,2} 
&=\frac{\mathcal{N}}{2\pi \hbar} \int_{-\frac{\beta}{2}}^{\frac{\beta}{2}} d\tau_0\frac{1}{A_{\tau_0}} \int_{0}^{\beta} d\xi \frac{1}{A_{\xi}}e^{-\frac{1}{\hbar}S_{cl,2}[\xi]}\prod_{q=2}^{\infty} \sqrt{\frac{1}{\lambda^{(2)}_q}}\;,\label{Eq.onebounce}
\end{align}
where we used the fact that the second contributions of the Jacobian prefactors %$\left|\frac{dc_{0}(\tau_0)}{d\tau_0}\right|$ and $\left|\frac{dc_{1}(\xi)}{d\xi}\right|$ 
in Eqs.~(\ref{Eq.Jac1}) and (\ref{Eq.Jac2}) are linear in $c_q$ and therefore vanish in the integration. The partition function contribution $z_{L}^{(2)}$ still depends on the action on the classical path $S_{cl,2}[\xi]$, on the product of eigenvalues $\prod_{q=2}^{\infty} \sqrt{\frac{1}{\lambda^{(2)}_q}}$ and on the Jacobians. 
In the following  sections we will deal with these quantities.

\section{Calculation of the prefactor $R_B$}\label{App:R}

\subsection{Basic Formula}
The ratio of determinants $R_B(\gamma,\tau_p)=\prod_{q=2}^{\infty} \sqrt{\lambda_{q}^{(0)}/\lambda_q^{(2)}}$ for one bounce path can be calculated by using the expression
\begin{equation}
\left(\prod_{q=2}^{\infty}{\lambda_q} \right)^{-\frac{1}{2}}=e^{-\frac{1}{2}\sum\limits_{q=2}^{\infty}\ln(\lambda_q)} =e^{-\frac{1}{2}\int_{m\omega_{0}^{2}}^{\infty}d\lambda\ln\left(\lambda\right)\rho(\lambda)}\;, \label{Eq.Detexp}
\end{equation}
where $\rho(\lambda)=\sum_{q=0}^\infty \delta(\lambda_q^{(2)}-\lambda)$ is  the density of eigenvalues. Using this we express the ratio of determinants as 
\begin{align}
R_B=e^{\frac{1}{2}\int_{m\omega_{0}^{2}}^{\infty}d\lambda\ln\left(\lambda\right)(\rho_0(\lambda)-\rho(\lambda))}\;, \label{Eq.factorR}
\end{align}
where $\rho_0(\lambda)=\sum_{q=0}^\infty \delta(\lambda^{(0)}_q-\lambda)$ is the density of eigenvalues in the harmonic oscillator potential and $\lambda^{(0)}_q$ are the solutions of the differential Eq.~(\ref{Eq.diff}) with the constant potential part $\mathcal{V}^*=m\omega_0^2$. By transforming Eq.~(\ref{Eq.diff}) to Matsubara frequencies we find 
\begin{align}
\lambda_{q}^{(0)}&=m\omega_{l}^{2}+m\omega_{0}^{2}+F_{l}+\omega_l^2\widetilde{F}_l\;, 
\end{align}
where $F_l$ and $\widetilde{F}_l$ are the Matsubara transformed kernels for the two dissipative couplings respectively. 
Using this we find for the density of eigenvalues
\begin{align}
\rho_0 (\lambda)&=\frac{1}{\pi}\text{Im}\left(\frac{1}{\beta}\sum_{l=-\infty}^{\infty}\frac{1}{m\omega_{l}^{2}+m\omega_{0}^{2}+F_{l}+\omega_l^2\widetilde{F}_l-\lambda-i\epsilon}\right)\nonumber\\
&=\frac{1}{\pi}\text{Im}\left(\frac{1}{2\pi }\int_{-\infty}^\infty d\omega\;{\widetilde{G}_\lambda^{(0)}(\omega)}\right)\;,\label{Eq.harmgreen}
\end{align}
where we took the low temperature limit and transformed to continuous frequencies $\omega$ and $\epsilon\rightarrow0$. Further, we defined the Greens function in Matsubara space $\widetilde{G}_\lambda^{(0)}(\omega) $ with the Fourier transform to imaginary time
\begin{equation}
G_{\lambda}^{(0)}(\tau)=\frac{1}{2\pi m}\int_{-\infty}^{\infty}d\omega\; \widetilde{G}_\lambda^{(0)}(\omega)  e^{i\omega\tau} \label{Eq.zerogreen}\;.
\end{equation}
For the density of solutions for Eq.~(\ref{Eq.diff}) with $\mathcal{V}(\tau)=\frac{m\omega_{0}^{2}}{2}\left(2-4\delta(x^{(2)}_{cl}(\tau))a_{0}\right)$ we use analogously 
\begin{align}
\rho(\lambda)=\frac{1}{\pi}\text{Im}\left(\frac{1}{2\pi }\int_{-\infty}^\infty d\omega \;\widetilde{G}_\lambda(\omega,\omega)  \right)\;,\label{(Eq.densityofstates)}
\end{align}
where the Fourier transform 
\begin{equation}
G_{\lambda}(\tau,\tau'')=\iint_{-\infty}^{\infty}\frac{d\omega}{2\pi } d\omega''\widetilde{G}_{\lambda}(\omega,\omega'')e^{i\omega\tau}e^{-i\omega''\tau''}\;, \label{Eq.convention}
\end{equation}
can be calculated via the Lippmann Schwinger equation
\begin{widetext}
\begin{align}
G_{\lambda}(\tau,\tau'')&=G_{\lambda}^{(0)}(\tau-\tau'')+2m\omega_{0}^{2}a_{0}\int_{-\infty}^{\infty} d\tau'G_{\lambda}^{(0)}(\tau-\tau')\delta(x_{cl}(\tau'))G_{\lambda}(\tau'',\tau')\;. \label{Eq.Lippmanntau}
\end{align}
yielding the solution
\begin{align}
G_{\lambda}(\tau,\tau'')=G_{\lambda}^{(0)}(\tau-\tau'')&+U\left(G_{\lambda}^{(0)}(\tau+\frac{\xi}{2})\left(G_{\lambda}^{(0)}(\frac{\xi}{2}-\tau'')N_{\lambda,2}(\xi)+G_{\lambda}^{(0)}(\frac{\xi}{2}+\tau'')N_{\lambda,1}(\xi)\right)\right)\nonumber\\&+U\left(G_{\lambda}^{(0)}(\tau-\frac{\xi}{2})\left(G_{\lambda}^{(0)}(\frac{\xi}{2}-\tau'')N_{\lambda,1}(\xi)+G_{\lambda}^{(0)}(\frac{\xi}{2}+\tau'')N_{\lambda,2}(\xi)\right)\right)\;, \label{Eq.sollippmann}
\end{align}
\end{widetext}
with $ U=2m\omega_{0}^{2}a_{0}/ \left|   \dot{x}_{cl}^{(2)}(\frac{\xi}{2}) \right| $, where $ \dot{x}^{(2)}_{cl}(\frac{\xi}{2})$ is the derivative of the classical bounce path at $\tau=\xi/2$  and we defined
\begin{align}
N_{\lambda,1}(\xi)	
&=\frac{1}{U}\frac{\left[U^{-1}-G_{\lambda}^{(0)}(0)\right]}{n_{\lambda}^{+}(\xi) n_{\lambda}^{-}(\xi) }\\
N_{\lambda,2}(\xi)	
&=\frac{1}{U}\frac{G_{\lambda}^{(0)}(\xi)}{n_{\lambda}^{+}(\xi) n_{\lambda}^{-}(\xi) }\;,
\end{align}
with $ n^{\pm}_\lambda(\xi) =  U^{-1}-G_{\lambda}^{(0)}(0)\pm G_{\lambda}^{(0)}(\xi)$.
We calculate the density of eigenvalues $\rho(\lambda)$, by Fourier transforming Eq.~(\ref{Eq.sollippmann}) back to the frequencies $\omega$. 
We obtain the form
\begin{align}
\widetilde{G}_{\lambda}(\omega,\omega')
&=\widetilde{G}_{\lambda}^{(0)}(\omega)\delta(\omega-\omega')\nonumber\\&+\widetilde{G}_{\lambda}^{(0)}(\omega)T_{\lambda}(\omega,\omega')\widetilde{G}_{\lambda}^{(0)}(\omega')\;, \label{Eq.finalgreen}
\end{align}
where we defined the $T$-matrix
\begin{align}
T_\lambda(\omega,\omega')&=\frac{2\cos\left((\omega+\omega')\frac{\xi}{2}\right){G}_{\lambda}^{(0)}(\xi)}{n_{\lambda}^{+}(\xi)n_{\lambda}^{-}(\xi)}\nonumber\\&+\frac{2\cos\left((\omega-\omega')\frac{\xi}{2}\right)\left[U^{-1}-{G}_{\lambda}^{(0)}(0)\right]}{n_{\lambda}^{+}(\xi)n_{\lambda}^{-}(\xi)}\;.
\end{align}
With Eq.~(\ref{Eq.finalgreen}) we can rewrite Eq.~(\ref{Eq.factorR}) as a function only depending on ${G}_{\lambda}^{(0)}(\xi)$. We find for the difference between the densities of eigenvalues in Eq.~(\ref{Eq.factorR})
\begin{align}
\rho_0(\lambda)-\rho(\lambda)
&=-\frac{1}{2\pi}\int_{-\infty}^\infty d\omega\; \widetilde{G}_{\lambda}^{(0)}(\omega)T_{\lambda}(\omega,\omega)\widetilde{G}_{\lambda}^{(0)}(\omega)\;.\label{Eq.factorR2}
\end{align} 
After some calculation we identify the equality 
\begin{align}
\frac{d}{d\lambda}\ln&\left(n_{\lambda}^{+}(\xi)n_{\lambda}^{-}(\xi)\right) =\nonumber \\ &-\frac{1}{2\pi}\int_{-\infty}^\infty d\omega\; \widetilde{G}_{\lambda}^{(0)}(\omega)T_{\lambda}(\omega,\omega)\widetilde{G}_{\lambda}^{(0)}(\omega)
\end{align}
and use the relation $ \text{Im}\left(\ln(x)\right)=\arg(x) $ to define the phases
\begin{align}
\phi^{(\pm)}_\lambda = \text{arg}\left(  U^{-1}-G_{\lambda}^{(0)}(0)\pm G_{\lambda}^{(0)}(\xi)\right) \label{Eq.phases}\;,
\end{align}
with the factor $U^{-1}$ from below Eq.~(\ref{Eq.sollippmann}).
With this we perform the partial integration with respect to $\lambda$ in the exponent of Eq.~(\ref{Eq.factorR}) yielding 
\begin{align}
\ln(R_B)&=\frac{1}{2\pi}\left[\ln\left(\lambda\right)\left(\phi^{+}_\lambda+\phi^{-}_\lambda\right)\right]_{m\omega_{0}^{2}}^{\infty} \nonumber\\ &-\frac{1}{2\pi}\int_{m\omega_{0}^{2}}^{\infty}d\lambda\frac{1}{\lambda}\left(\phi^{+}_\lambda+\phi^{-}_\lambda\right)\;. \label{Eq.lnR}
\end{align}
The phases satisfy $\phi_{m\omega_0^2}^{\pm}=-\pi$ (calculated in the next section) and $\lim\limits_{\lambda\rightarrow \infty}\phi_\lambda^{\pm}=0$ leading to the result
\begin{align}
\ln(R_B)&=\ln\left(m\omega_0^2\right) -\frac{1}{2\pi}\int_{m\omega_{0}^{2}}^{\infty}d\lambda\frac{1}{\lambda}\left(\phi^{+}_\lambda+\phi^{-}_\lambda\right)\;. \label{Eq.lnRfinal}
\end{align}
The last step is to calculate the phases $\phi_{\lambda}^{\pm}$ depending on the Greens function $G_{\lambda}^{(0)}(\tau)$ defined in Eq.~(\ref{Eq.zerogreen}).

\subsection{Calculation of the phases $\phi_{\lambda}^{\pm}$ }
In the last subsection we defined the quantities $\phi_{\lambda}^{\pm}$ in Eq.~(\ref{Eq.phases}).
We start by calculating the Greens function $G_{\lambda}^{(0)}(\tau)$. In presence of ohmic unconventional dissipation this quantity diverges, because of diverging position quantum fluctuations. This divergence is well known in the literature and can be cured by adding a Drude cutoff function $f_c(\omega)$ with high frequency cutoff $\omega_c$. The dissipative kernel for unconventional dissipation then reads 
\begin{equation}
 \widetilde{F}(\omega)=\frac{-\tau_{p}|\omega|m f_c(\omega)}{1+\tau_{p}|\omega|f_c(\omega)}\;,
\end{equation}
with $f_c(\omega)=(1+|\omega|/\omega_c)^{-1}$. We also insert the dissipative kernel for conventional dissipation, defined in the main text, into  Eq.~(\ref{Eq.zerogreen}) and find 
\begin{align}
G_{p}^{(0)}(\tau)\label{Eq. Fourierintegral}
&=\\&\frac{1}{\pi m\omega_{c}}\int_{0}^{\infty}dx\frac{\left(1+(1+\tau_{p}\omega_{c})x\right)\cos(x\omega_{c}\tau)}{-p^{2}\Omega_{c}^{2}+\chi(p^2) x+\alpha x^{2}+x^{3}-i\epsilon^*}\nonumber\;, 
\end{align}
with $p^2=\frac{\lambda}{m\omega_0^2}-1$,  $x=\omega/\omega_c$, $\Omega_c=\omega_0/\omega_c$, $\alpha=\left({\gamma}/{\omega_{c}}+\tau_{p}\gamma+1\right)$ and 
\begin{align}
\chi(p^2)&=\left(-p^{2}\Omega_{c}^{2}(1+\tau_{p}\omega_{c})+\frac{\gamma}{\omega_{c}}\right)\;.
\end{align}
The denominator of Eq.~(\ref{Eq. Fourierintegral}) is a cubic polynomial with an imaginary part. We expand the polynomial part into its roots $\tilde{x}_{1,2,3}$ and obtain by introducing $\tilde{x}_{1}=\nu_1$, $\tilde{x}_{2}=-\nu_2$ and $\tilde{x}_{3}=-\nu_3$
\begin{align}
-p^{2}\Omega_{c}^{2}-\chi(p^2)& x+\alpha  \label{Eq.greenpol} x^{2}+x^{3}\\&=(x-\nu_1)(x+\nu_2)(x+\nu_3)\;,\nonumber
\end{align}
where $\nu_i>0$. Because the full form of the quantities $\nu_{1,2,3}$ is not important at this stage, we do not present them here explicitly, but refer to App.~\ref{Sec.Aux}. With this definition we can perform a principle value integration in Eq.~(\ref{Eq. Fourierintegral}). We obtain the result
\begin{align}
G_p^{(0)}(\tau)
&=\frac{1}{\pi m\omega_{c}}\left[\sum_{i=1}^{3}\widetilde{\mathcal{U}}_ig[\tau\omega_c\nu_i]-\widetilde{\mathcal{U}}_1\pi\sin(\tau\omega_c\nu_1)\right]\nonumber\\&+\frac{i}{ m\omega_{c}}\frac{\left(1+(1+\tau_{p}\omega_{c})\nu_{1}\right)\cos(\nu_{1}\omega_{c}\tau)}{|\chi(p^{2})+2\alpha\nu_{1}+3\nu_{1}^{2}|}\;,\label{Eq.finalgreentau}
\end{align}
where  the prefactors $\widetilde{\mathcal{U}}_i$ originate from a partial fraction expansion  also defined in App.~\ref{Sec.Aux}. The functions $g(x)$ are defined in App.~\ref{Sec.Fullaction}. For the factor  $U^{-1}$ we have to calculate
\begin{align}
\dot{x}^{(2)}_{cl}\left(\frac{\xi}{2}\right)&=\frac{2\omega_{0}^{2}a_{0}}{\pi}\int_{0}^{\infty}dx\frac{\left(1+(1+\tau_{p}\omega_{c})x\right)\left[1-\cos(\xi\omega)\right]}{\Omega_{c}^{2}+\chi(1) x+\alpha x^{2}+x^{3}}\;, \label{Eq.U^-1}
\end{align}
which has no imaginary part. We calculate the integral by defining the polynomial in the denominator like
\begin{align}
\Omega_{c}^{2}+\chi(1) & x+\alpha   \nonumber x^{2}+x^{3}\\&=(x+k_1)(x+k_2)(x+k_3). \label{Eq.upol}
\end{align}
with $k_i>0$ and find 
\begin{align}
U^{-1}=-\frac{1}{\pi m\omega_c}\sum_{i=1}^{3}\widetilde{\mathcal{T}}_i\left(\ln\left(k_i\right)+g[k_i\omega_c\xi]\right)\;,
\end{align}
where the prefactors $\widetilde{\mathcal{T}}_i$ and the quantities $k_i$ are also defined in App.~\ref{Sec.Aux}.
Inserting $\tau=\xi$ in Eq.~(\ref{Eq.finalgreentau}) and calculating $G_p^{(0)}(\tau=0)$ we find the result 
\begin{align}
n_\lambda^{(\pm)}&=-\frac{1}{\pi m\omega_c}\sum_{i=1}^{3}\widetilde{\mathcal{T}}_i\left(\ln\left(k_i\right)+g[k_i\omega_c\xi]\right)\nonumber\\&+\frac{1}{\pi m\omega_{c}}\left[\sum_{i=1}^{3}\widetilde{\mathcal{U}}_i\left(\ln(\nu_{i})\pm g[\xi\omega_c\nu_i]\right)\mp\widetilde{\mathcal{U}}_1\pi\sin(\xi\omega_c\nu_1)\right]\nonumber
\\&-\frac{i}{ m\omega_{c}}\frac{\left(1+(1+\tau_{p}\omega_{c})\nu_{1}\right)\left(1\mp\cos(\nu_{1}\omega_{c}\xi)\right)}{|\chi(p^{2})+2\alpha\nu_{1}+3\nu_{1}^{2}|}\;. \label{Eq.npm}
\end{align}
We calculate the phases via $\phi_{\lambda}^{(\pm)}=\text{arg}(n_\lambda^{(\pm)})$. In the last section we used $\phi_{m\omega_0^2}^{(\pm)}=-\pi$. We briefly comment on this here.  We see from Eq.~(\ref{Eq. Fourierintegral}) that there is an infrared divergence for $G_p^{(0)}(0)$ by inserting $\lambda=m\omega_0^2\rightarrow p=0$. This means that $n_{m\omega^2_0}^{(\pm)}\rightarrow -\infty+i c$, where $c$ is a finite constant and therefore $\phi_{m\omega_0^2}^{(\pm)}\rightarrow-\pi$. 

\subsection{Final result for $R_B$}
We obtain the final solution for the ratio of determinants by inserting $n_\lambda^{(\pm)}(\xi)$ into Eq.~(\ref{Eq.lnRfinal}). We  substitute $p^2=\frac{\lambda}{m\omega_0^2}-1$ and obtain
\begin{equation}
\ln(R_B)=\ln\left(m\omega_0^2\right)-\frac{1}{\pi }\int_{0}^{\infty}dp\frac{p}{1+p^2}\left(\phi_p^{(+)}+\phi_p^{(-)}\right)\;. \label{Eq.numint}
\end{equation}
In the limit of a large bounce $\xi\omega_c\nu_i\gg1$ the function $g[\xi\omega_c\nu_i]\propto 1/\xi^2$ and can therefore be neglected (see App.~\ref{App.factor}). The quantity $\nu_1$ depends on $p$ (see. App.~\ref{Sec.Aux}). In the limit $\xi\omega_c\nu_1\gg1$ the functions $\sin(\xi\omega_c\nu_1)$ and $\cos(\xi\omega_c\nu_1)$ in Eq.~(\ref{Eq.npm}) are highly oscillating. They do only weakly contribute to the integral over $p$ and we neglect them. With this approximations we find
\begin{equation}
\ln(R_B)=\ln\left(m\omega_0^2\right)-\frac{2}{\pi }\int_{0}^{\infty}dp\frac{p}{1+p^2}\arg\left(U^{-1}-G_{p}^{(0)}(0)\right) \label{Eq.numint2}\;,
\end{equation}
which we solve numerically leading to the result shown in Fig.~\ref{Fig.Unconresults}b and Fig.~\ref{Fig:phasechange2}b. 

\section{The Jacobian factor $\mathcal{L}_B$}\label{App.L}
Here we show the calculation for Jacobian prefactor for one bounce path. As defined above the factor due to the change of the integral variables reads
\begin{align}
\mathcal{L}_B(\gamma,\tau_{p})&=\frac{1}{A_{\tau_0}}\frac{1}{A_{\xi}}\nonumber\\&=\sqrt{\int_{-\beta/2}^{{\beta/2}}d\tau\dot{x}_{cl,2}^{2}(\tau,\xi)}\sqrt{\int_{-\beta/2}^{{\beta/2}}d\tau x_{cl,2}'^{2}(\tau,\xi)}\;, \label{Eq.factorL}
\end{align}
where $ \dot{x}_{cl,2}=\frac{\partial{x}_{cl,2}(\tau,\xi)}{\partial\tau} $ and $ {x'}_{cl,2}=\frac{\partial{x}_{cl,2}(\tau,\xi)}{\partial\xi} $ with $x_{cl,2}$ defined in Eq.~(\ref{Eq.bounce}) and Eq.~(\ref{Eq.velocityclasscialpathmatsu}).
The derivative with respect to $ \tau $ yields
\begin{align}
\dot{x}_{cl,2}=\frac{4\omega_{0}^{2}a_{0}}{\beta}\sum_{l=1}^{\infty}v^{(cl)}_{l}\left[\cos(\omega_{l}(\tau+\frac{\xi}{2}))-\cos(\omega(\tau-\frac{\xi}{2}))\right]\;.
\end{align}
Inserting $ \dot{x}^2_{cl,2} $ in to the first factor of Eq. (\ref{Eq.factorL}), performing the integration over $ \tau  $ and substituting the sum with an integral in the small temperature limit we obtain the integral over frequencies $ \omega $
\begin{align}
\int_{-\frac{\beta}{2}}^{\frac{\beta}{2}}d\tau\dot{x}_{cl,2}^{2}(\tau,\xi)=\frac{8\omega_{0}^{4}a_{0}^{2}}{\pi}\int_{0}^{\infty}d\omega\frac{(1+\tau_{p}\omega)^{2}(1-\cos\left(\xi\omega\right))}{(\omega_{0}^{2}+\bar{\Gamma}\omega+\sigma\omega^{2})^{2}}\;,\label{Eq.veloL}
\end{align} 
where $ \bar{\Gamma}=\tau_{p}\omega_{0}^{2}+\gamma $ and $ \sigma=1+\tau_{p}\gamma$.
The integration over the summand containing $ \cos(\xi\omega) $ only leads to corrections $ O(1/\xi^2) $ which we neglect in the limit $\xi\omega_0\gg1$. The rest of the integral yields the $ \xi  $ independent part 
\begin{align}
\int_{-\frac{\beta}{2}}^{\frac{\beta}{2}}d\tau\dot{x}_{cl,2}^{2}(\tau,\xi)=\frac{8\omega_{0}^{2}a_{0}^{2}}{\pi} \mathcal{C}(\gamma,\tau_p)\;, 
\end{align} 
where we introduced 
\begin{align}
\mathcal{C}(\gamma,\tau_p)&=\frac{\frac{2\Gamma_+}{\omega_{0}}-4\tau_{p}+\frac{2\tau_{p}^{2}\omega_{0}\Gamma_+}{\sigma}}{(4\Gamma_+^{2}-4\sigma)}\\\nonumber&+\left(1-\frac{2\tau_{p}\omega_{0}\Gamma_+}{\sigma}+\frac{\tau_{p}^{2}\omega_{0}^{2}}{\sigma}\right)\frac{2\sigma\ln\left(\frac{\Lambda_{2}}{\Lambda_{1}}\right)}{\omega_{0}\left(\sqrt{4\Gamma_+^{2}-4\sigma}\right)^{3}}\;,
\end{align}
with and $\Lambda_{1,2}$ defined in Eq.~(\ref{Eq.roots})
Using the same procedure we find for the "breathing mode velocity" 
\begin{align}
{x'}_{cl,2}(\tau,\xi)&=v_{0}\nonumber\\&+\sum_{l=1}^{\infty}\frac{v^{(cl)}_{l}}{2}\left[\cos(\omega_{l}(\tau+\frac{\xi}{2}))+\cos(\omega_{l}(\tau-\frac{\xi}{2}))\right]\;.
\end{align}
Analogously to the translational velocity we get
\begin{align}
\int_{-\beta/2}^{{\beta/2}}d\tau\nonumber x_{cl,2}'^{2}(\tau,\xi)=\frac{2\omega_{0}^{2}a_{0}^{2}}{\pi}\mathcal{C}(\gamma,\tau_p)\;.
\end{align} 
Therefore we finally find
\begin{align}
\mathcal{L}_B(\gamma,\tau_{p})&=2\frac{V_{0}}{m\omega_{0}}\mathcal{P}(\gamma,\tau_{p})\;,\label{Eq.Pinal}
\end{align}
where
\begin{align}
\mathcal{P}(\gamma,\tau_{p})=\frac{4}{\pi}\mathcal{C}(\gamma,\tau_p)\;.
\end{align}
In the non dissipative case we recover $ \mathcal{P}(0,0) =1 $. The behavior of $ \mathcal{L}_B(\gamma,\tau_p) $ is shown in Figs.~\ref{Fig.Unconresults}c and \ref{Fig:phasechange2}c.

\begin{figure}[t]
	\centering
	\includegraphics[width=1.0\linewidth]{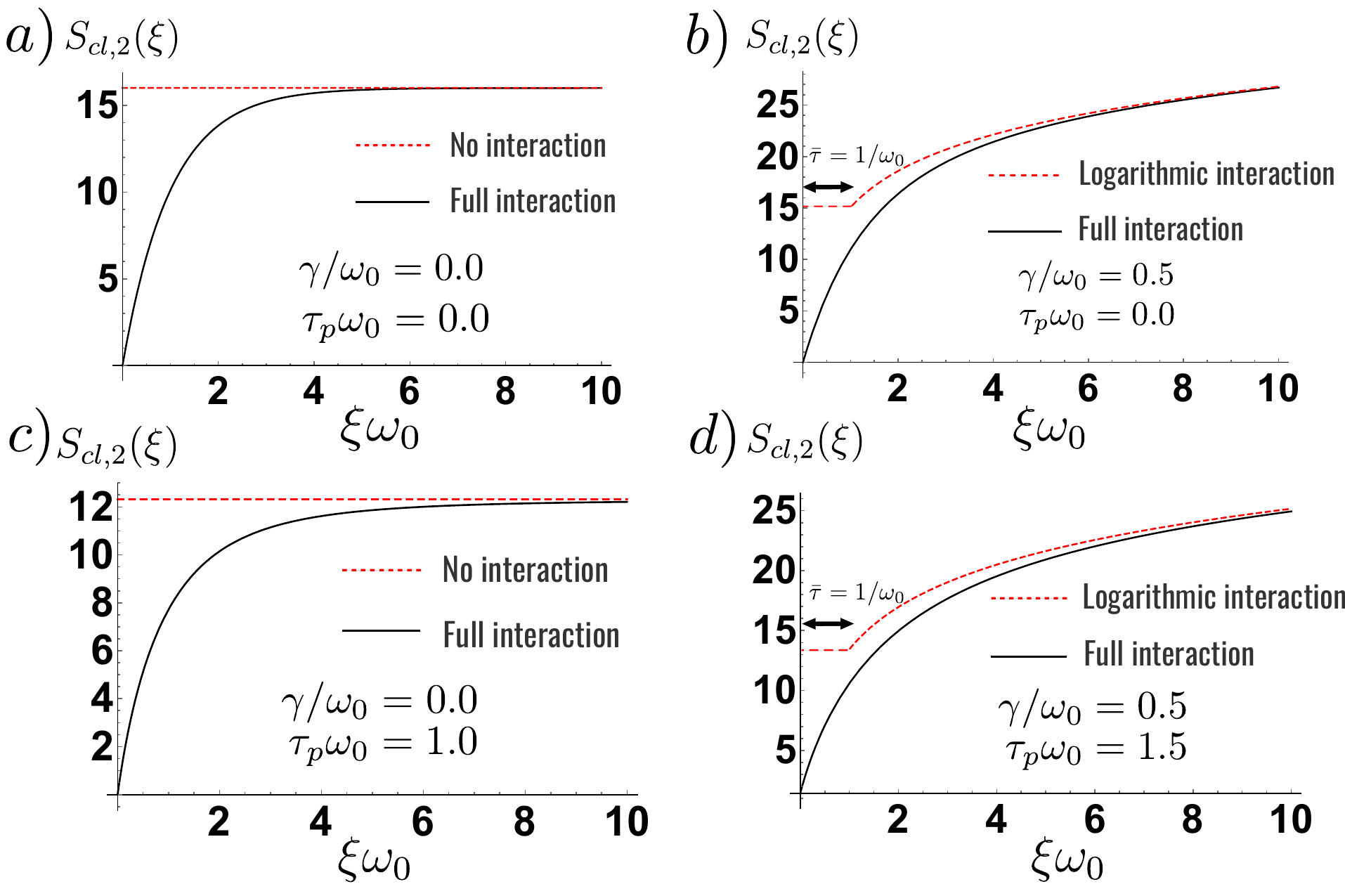}

	\caption{Action on the classical path as a function of $\xi$ for different dissipative cases: a)  No dissipation. The action becomes zero for vanishing distance. The instantons decouple exponentially for large $\xi$. b) Conventional dissipation: Logarithmic long distance behavior. c) Unconventional dissipation: The instantons decouple with $1/(\xi\omega_0)^2$. d) Frustrated dissipation: same long range behavior as in the conventional dissipative case.}
	\label{Fig.interaction}
\end{figure}

\section{Extended bounce approximation}\label{App.dilute}
\subsection{Action on the classical path for frustrated dissipation} \label{Sec.Fullaction}
To calculate the action on the classical path for a bounce path, we use the ansatz Eq.~(\ref{Eq.bounce}) yielding the intergral in Eq.~(\ref{Eq:action2}). Solving this integral we find the chemical potential part in Eq.~(\ref{Eq.fullchemical}). Further, due to the cosine in (\ref{Eq:action2}) we find the interaction part

\begin{align}
\mathcal{I}(\gamma,\tau_p,\xi)& =\frac{\eta}{\omega_{0}}\left(\ln(\omega_{0}\xi)+\frac{1}{2}\left(g(\xi\Lambda_{1})+g(\xi\Lambda_{2})\right)\right)\nonumber\\&-\frac{8V_0}{\pi\hbar\omega_0}\frac{\left(\frac{\gamma}{\omega_{0}}\Gamma_{-}-1\right)}{2\sqrt{\Gamma_{-}^{2}-1}}\left[g(\xi\Lambda_{1})-g(\xi\Lambda_{2})\right]\label{eq:finalaction}\;,
\end{align}
where the quantities $\Lambda_{1,2}$ are defined in Eq.~(\ref{Eq.roots}) with the functions
\begin{align}
g(ax)=&-\cos(ax)Ci(ax)-\sin(ax)Si(ax)+\frac{1}{2}\sin(ax)\pi \nonumber\\
&=\int_{0}^{\infty}\frac{\cos\left(a\omega\right)}{(x+\omega)}d\omega\;.
\end{align}
For $\xi\omega_0 \gg1$ and $\Gamma_-<1$ we find that $g(\Lambda_{1}\xi)\pm g(\Lambda_{2}\xi) \propto 1/(\xi\omega)^2$. The short distance behavior of the functions $g(ax)$ is responsible for the annihilation process we show in Fig.~\ref{Fig.bouncepathdiss}.  If the distance $\xi\omega_0$ is large enough to neglect this function we call the resulting two instanton trajectory an \textit{extended} bounce. In the main text we only consider instantons that reach far into the right parabolic well as they contribute to the delocalization transition.This  \textit{extended} configurations are manifested via an instanton hard core size $\bar{\tau}$. Hence, the \text{extended} bounces we consider are non interacting in the non dissipative and in the unconventional, while logarithmically interacting in the conventional and in the frustrated dissipative case. 
In the discussion of the main text we choose $\bar{\tau}=1/\omega_0$ for the small distance cutoff. Using this, the lowest energy \textit{extended} bounce path has spacing $\xi\omega_0\approx 1$ (see. Fig.~\ref{Fig.interaction}b and d) and is logarithmically interacting (in the conventional and frustrated case).

%Note, that at this distance the $g(x)$-function are already important and the bounce is partly annihilated making the cutoff path artificial.  However, if this artificial path is not possible in the Anderson renormalization treatment the more extended configurations are also not possible as they have higher energy. Hence, strictly speaking the cutoff $\bar{\tau}=1/\omega_0$ is too small, but we are on the safe side of the approximation and the qualitative nature of phase diagram does not change by choosing a higher cutoff. 

%Further, in the case of a short bounce, we find that the action on the classical path $S_{cl,2}\rightarrow 0 $ as the distance approaches zero. The $\xi$ dependence of the interaction is shown in Fig.~\ref{Fig.interaction}.

\begin{figure}[t]
	\centering
	\includegraphics[width=0.49\linewidth]{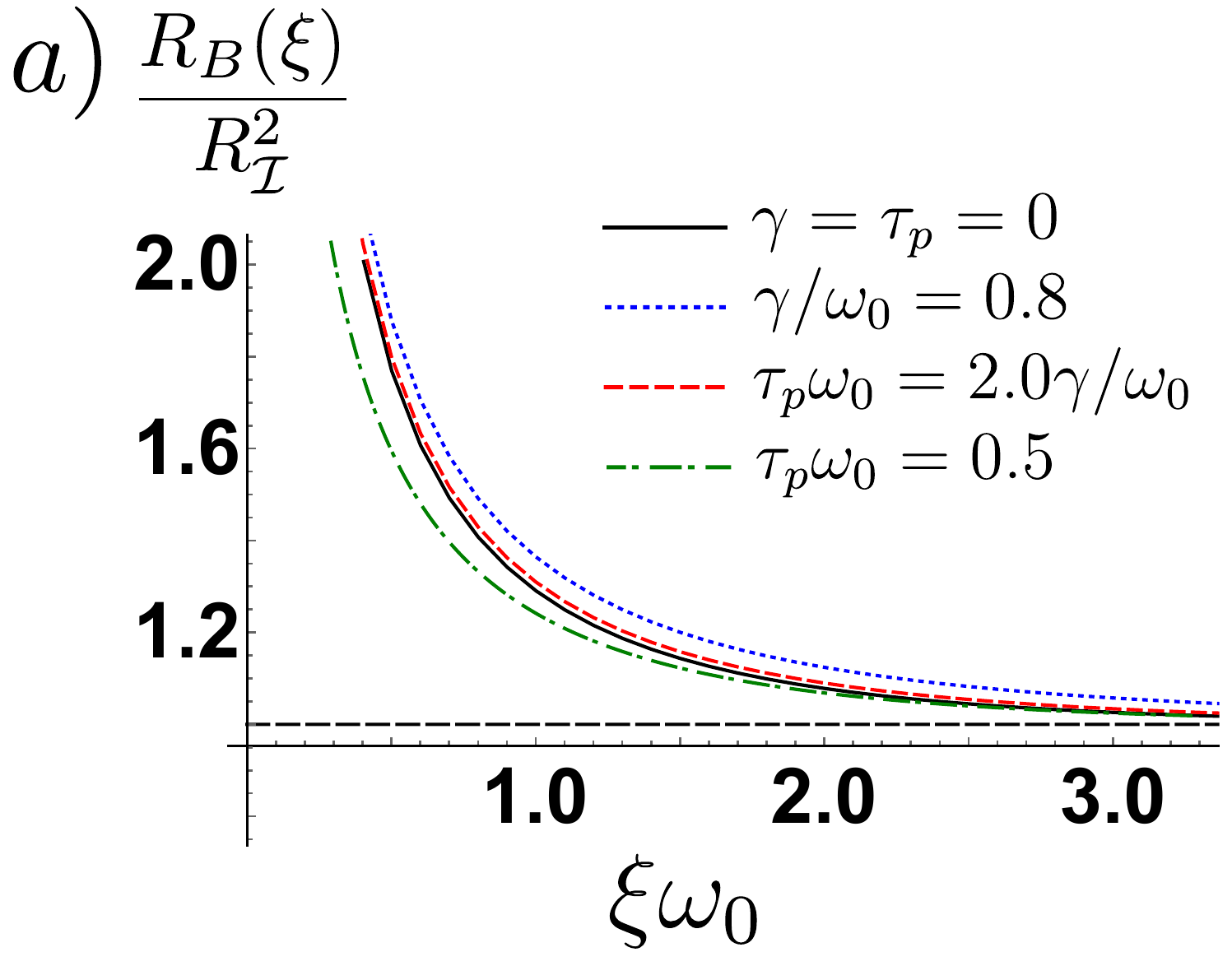}
	\includegraphics[width=0.49\linewidth]{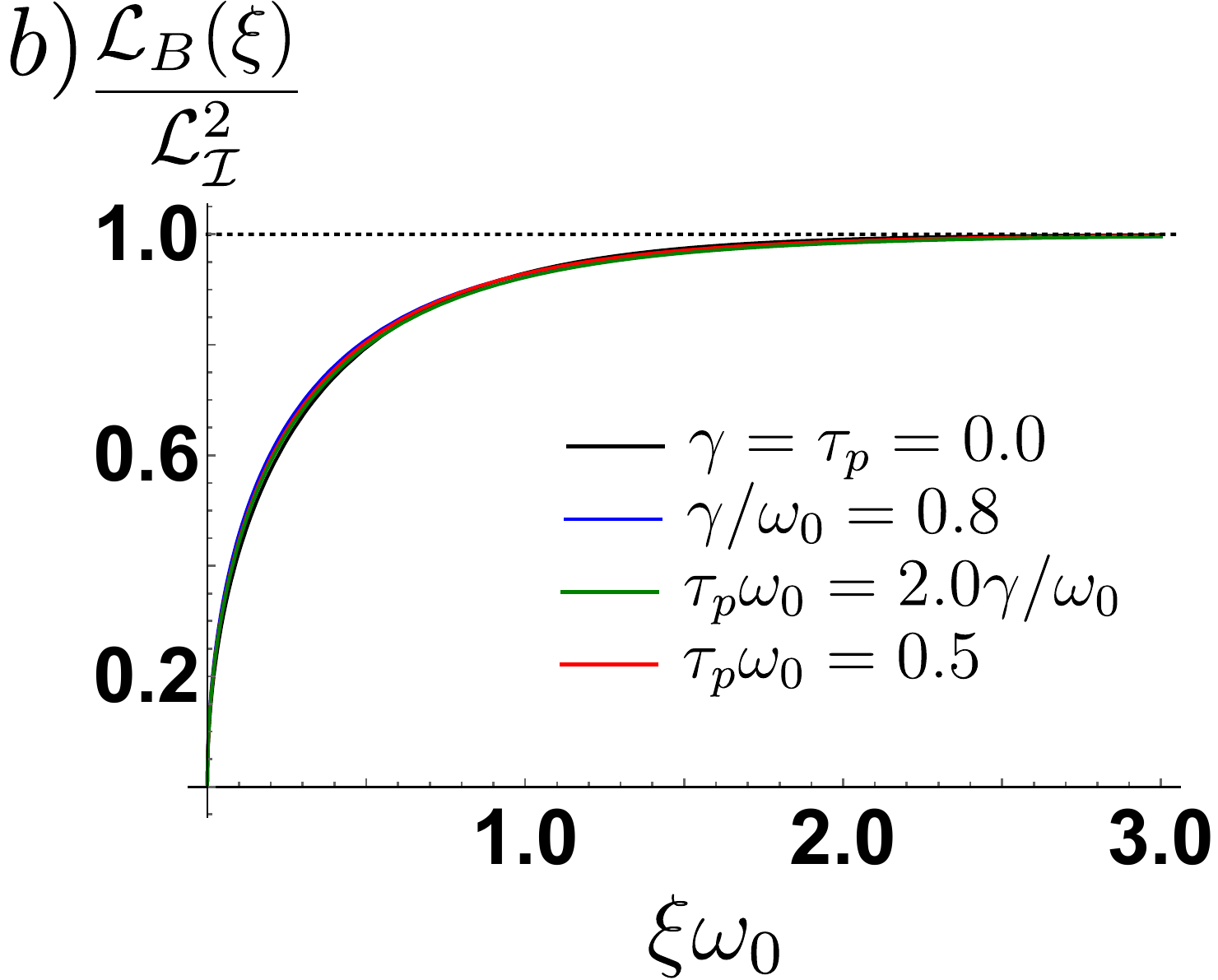}\\
		\includegraphics[width=1.0\linewidth]{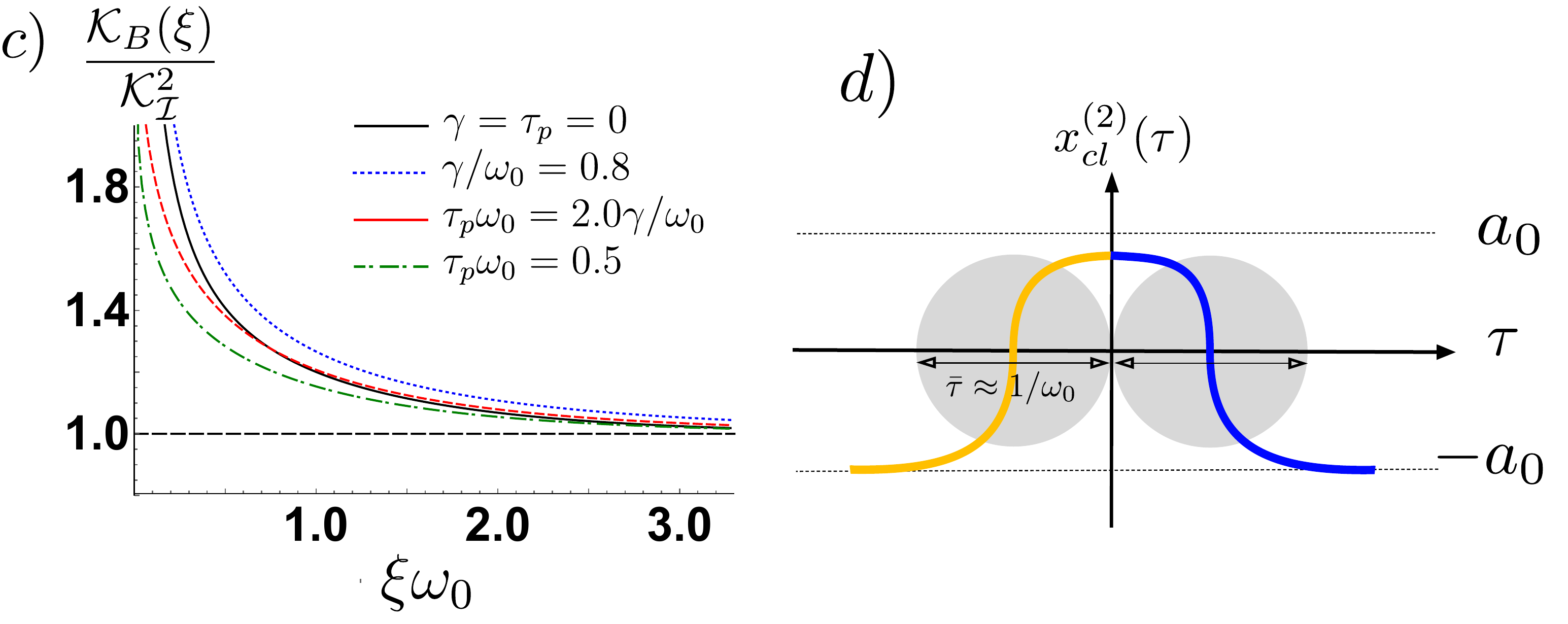}
	\caption{a)-c) $\xi$-dependence of the fluctuations prefactors and $\mathcal{K}_B(\xi)=R_B(\xi)\mathcal{L}_B(\xi)$, scaled with the square of the respective quantities for one instanton $\mathcal{K}^2_\mathcal{I}=R^2_\mathcal{I}\mathcal{L}^2_\mathcal{I}$ . d) System used in the discussion of the paper. } 
	\label{Fig:Rxi}
\end{figure}
\subsection{Factorization of the activity}\label{App.factor}
We here show that the factorization of the activity $\mathcal{A}_{kB}=\mathcal{A}_\mathcal{I}^{2k}$ for $k$-bounces, defined in Eq.~(\ref{Eq.activity1}) and Eq.~(\ref{Eq.activity2}) is a good approximation in the limit $\xi\omega_0\gg1$.
%The quantity $\epsilon_0$ does not depend on the distance between the instantons by construction. The production of $k$-bounces yields the chemical potential $k \mu$.
The activity in Eq.~(\ref{Eq.activity1}) contains the quantities  $\mathcal{L}_B $ and $R_B$ originating from the fluctuation prefactor. During the calculation we use $R_B=R_\mathcal{I}^2$ and $\mathcal{L}_B=\mathcal{L}_\mathcal{I}^2$, where $R_\mathcal{I}$ and $\mathcal{L}_\mathcal{I}$ are the respective quantities for one instanton. Here we show  this valid by analyzing $\mathcal{K}_B(\xi)=\mathcal{L}_B(\xi) R_B(\xi)$. If the $\xi$-dependence vanishes, the factors of each instanton become independent of each other and the bounce prefactor factorizes. 
We start with the ratio of determinants $R_B$ defined in Eq.~(\ref{Eq.lnRfinal}) with the $\xi$ dependent phases $\phi_\lambda^{(\pm)}$ defined in Eq.~(\ref{Eq.phases}) reading
\begin{align}
\phi^{(\pm)}_\lambda = \text{arg}\left(  U^{-1}-G_{\lambda}^{(0)}(0)\pm G_{\lambda}^{(0)}(\xi)\right)\;. 
\end{align}
We calculate the ratio  $R_B/R_\mathcal{I}^2$, and show in Fig.~\ref{Fig:Rxi}a that already for $\xi\omega_0\approx 3$  the values of $R_B(\xi)$ has almost converged to the $\xi$ independent one $R^2_\mathcal{I}$ . As we analyze the system in the limit  $\xi\omega_0\gg1$ we neglect the $\xi$-dependence and use the factorization $R_{kB}=R_\mathcal{I}^{2k}$ manifested in the choice of 
\begin{align}
\phi_\lambda^{(\pm)}= \text{arg}\left(  U^{-1}-G_{\lambda}^{(0)}(0)\right)\;,
\end{align}
we use in Eq.~(\ref{Eq.numint2}).
% In the above treatment we still neglected the oscillatory terms in $G_\lambda^{(0)}(\xi)$ as they only weakly contribute to the integral over $p$. 
The $\xi$-dependence of the quantity $\mathcal{L}_B$ originates from the second term in Eq.~(\ref{Eq.veloL}). For $\xi\rightarrow0$ we see that the factor will vanish because $\cos(0)=1$ and the integrand is zero. In Fig.~\ref{Fig:Rxi}b we further see how $\mathcal{L}_B(\xi)$ converges to $\mathcal{L}_\mathcal{I}^2$. 
We also show the full fluctuation prefactor $\mathcal{K}_B(\xi)=R_B(\xi)\mathcal{L}_B(\xi)$ in Fig.~\ref{Fig:Rxi}c and its $\xi$-dependence. We see that it converges quickly to the value used in the main text. 
Finally, because we choose the quantity $\epsilon_0$ to be the part of the action that is independent of $\xi$ it satisfies $\epsilon_0^{(k)}=k\epsilon_0$ by construction. This leads to the factorization of the full bounce activity defined in Eq.~(\ref{Eq.activity1}) yielding Eq.~(\ref{Eq.activity2}).

\vspace{1cm}
\section{Auxiliary Variables }\label{Sec.Aux}
In the above section we introduced the quantities  $\widetilde{\mathcal{T}}_i$, $\widetilde{\mathcal{U}}_i$, $\nu_i$, and $k_i$. The first two originate from the partial fraction expansions of the integrands in Eq.~(\ref{Eq. Fourierintegral}) and Eq.~(\ref{Eq.U^-1}).  The latter ones are related to the roots of the denominators of $\widetilde{G}^{(0)}_\lambda(\omega)$ and $U^{-1}$. The prefactors read
\begin{align}
\widetilde{\mathcal{T}}_1&=\frac{1-(1+\tau_p \omega_c){k}_1}{({k}_1-{k}_2)({k}_1-{k}_3)}\\ \widetilde{\mathcal{T}}_2&=\frac{-1+(1+\tau_p \omega_c){k}_2}{({k}_1-{k}_2)({k}_2-{k}_3)}\\ \widetilde{\mathcal{T}}_3&=\frac{-1+(1+\tau_p \omega_c)k_3}{({k}_1-{k}_3)({k}_3-\bar{k}_2)}
\end{align}
and
\begin{align}
\widetilde{\mathcal{U}}_1&=\frac{1+(1+\tau_p \omega_c){\nu}_1}{(\nu_1+{\nu}_2)({\nu}_1+{\nu}_3)}\\ \widetilde{\mathcal{U}}_2&=\frac{1-(1+\tau_p \omega_c){\nu}_2}{({\nu}_1+{\nu}_2)({\nu}_2-{\nu}_3)}\\ \widetilde{\mathcal{U}}_3&=\frac{1-(1+\tau_p \omega_c)\nu_3}{(\nu_1+{\nu}_3)({\nu}_3-{\nu}_2)}\;.
\end{align}
\begin{widetext}
We use the basic formula for the roots of cubic polynomials and find for the roots in Eq.~(\ref{Eq.upol})
\begin{align}
-k_1&=\frac{\alpha}{3}+\frac{2^{\frac{1}{3}}\eta(1)}{3\left(\Sigma(1)+\sqrt{4\eta^{3}(1)+\Sigma^{2}}\right)^{\frac{1}{3}}}-\frac{\left(\Sigma(1)+\sqrt{4\eta^{3}(1)+\Sigma^{2}(1)}\right)^{\frac{1}{3}}}{3\cdot2^{\frac{1}{3}}}
\end{align}
\begin{align}
-k_2&=\frac{\alpha}{3}-\frac{(1+i\sqrt{3})\eta(1)}{3\cdot2^{\frac{2}{3}}\left(\Sigma(1)+\sqrt{4\eta^{3}(1)+\Sigma^{2}(1)}\right)^{\frac{1}{3}}}+\frac{(1-i\sqrt{3})\left(\Sigma(1)+\sqrt{4\eta(1)+\Sigma^{2}(1)}\right)^{\frac{1}{3}}}{6\cdot2^{\frac{1}{3}}}\\
-k_3&=\frac{\alpha}{3}-\frac{(1-i\sqrt{3})\eta(1)}{3\cdot2^{\frac{2}{3}}\left(\Sigma(1)+\sqrt{4\eta^{3}(1)+\Sigma^{2}(1)}\right)^{\frac{1}{3}}}+\frac{(1+i\sqrt{3})\left(\Sigma(1)+\sqrt{4\eta(1)+\Sigma^{2}(1)}\right)^{\frac{1}{3}}}{6\cdot2^{\frac{1}{3}}}\;,
\end{align}
where we defined
\begin{align}
\eta(1)&=(-\alpha^{2}-3\chi(1))\\
\Sigma(1)&=-2\alpha^{3}+27\Omega_{c}^{2}-9\alpha\chi(1)\;.
\end{align}
Further, the roots for the polynomial Eq. (\ref{Eq.greenpol}) read
\begin{align}
\nu_1&=\frac{\alpha}{3}+\frac{2^{\frac{1}{3}}\eta(p^2)}{3\left(\Sigma(p^{2})+\sqrt{4\eta^{3}(p^{2})+\Sigma^{2}(p^{2})}\right)^{\frac{1}{3}}}-\frac{\left(\Sigma(p^{2})+\sqrt{4\eta^{3}(p^{2})+\Sigma^{2}(p^{2})}\right)^{\frac{1}{3}}}{3\cdot2^{\frac{1}{3}}}\\
-\nu_2&=\frac{\alpha}{3}-\frac{(1+i\sqrt{3})\eta(p^{2})}{3\cdot2^{\frac{2}{3}}\left(\Sigma(p^{2})+\sqrt{4\eta^{3}(p^{2})+\Sigma^{2}(p^{2})}\right)^{\frac{1}{3}}}+\frac{(1-i\sqrt{3})\left(\Sigma(p^{2})+\sqrt{4\eta^3(p^{2})+\Sigma^{2}(p^{2})}\right)^{\frac{1}{3}}}{6\cdot2^{\frac{1}{3}}}\\
-\nu_3&=\frac{\alpha}{3}-\frac{(1-i\sqrt{3})\eta(p^2)}{3\cdot2^{\frac{2}{3}}\left(\Sigma(p^{2})+\sqrt{4\eta^{3}(p^{2})+\Sigma^{2}(p^{2})}\right)^{\frac{1}{3}}}+\frac{(1+i\sqrt{3})\left(\Sigma(p^{2})+\sqrt{4\eta^3(p^{2})+\Sigma^{2}(p^{2})}\right)^{\frac{1}{3}}}{6\cdot2^{\frac{1}{3}}}\;,
\end{align}
where
\begin{align}
\eta(p^2)&=-\alpha^{2}-3\chi(p^{2})\\
\Sigma(p^2)&=-2\alpha^{3}+27p^{2}\Omega_{c}^{2}-9\alpha\chi(p^{2})\;.
\end{align}
\end{widetext}
In the App.~\ref{App:R} we defined
$p^2=\frac{\lambda}{m\omega_0^2}-1$, $\Omega_c=\omega_0/\omega_c$, $\alpha=\left({\gamma}/{\omega_{c}}+\tau_{p}\gamma+1\right)$.

\bibliographystyle{mybibstyle}

\bibliography{bibliographydouble}

%merlin.mbs apsrev4-1.bst 2010-07-25 4.21a (PWD, AO, DPC) hacked
%Control: key (0)
%Control: author (72) initials jnrlst
%Control: editor formatted (1) identically to author
%Control: production of article title (1) required
%Control: page (0) single
%Control: year (1) truncated
%Control: production of eprint (0) enabled
\begin{thebibliography}{36}%
\makeatletter
\providecommand \@ifxundefined [1]{%
 \@ifx{#1\undefined}
}%
\providecommand \@ifnum [1]{%
 \ifnum #1\expandafter \@firstoftwo
 \else \expandafter \@secondoftwo
 \fi
}%
\providecommand \@ifx [1]{%
 \ifx #1\expandafter \@firstoftwo
 \else \expandafter \@secondoftwo
 \fi
}%
\providecommand \natexlab [1]{#1}%
\providecommand \emph  [1]{``#1''}%
\providecommand \bibnamefont  [1]{#1}%
\providecommand \bibfnamefont [1]{#1}%
\providecommand \citenamefont [1]{#1}%
\providecommand \href@noop [0]{\@secondoftwo}%
\providecommand \href [0]{\begingroup \@sanitize@url \@href}%
\providecommand \@href[1]{\@@startlink{#1}\@@href}%
\providecommand \@@href[1]{\endgroup#1\@@endlink}%
\providecommand \@sanitize@url [0]{\catcode `\\12\catcode `\$12\catcode
  `\&12\catcode `\#12\catcode `\^12\catcode `\_12\catcode `\%12\relax}%
\providecommand \@@startlink[1]{}%
\providecommand \@@endlink[0]{}%
\providecommand \url  [0]{\begingroup\@sanitize@url \@url }%
\providecommand \@url [1]{\endgroup\@href {#1}{\urlprefix }}%
\providecommand \urlprefix  [0]{URL }%
\providecommand \Eprint [0]{\href }%
\providecommand \doibase [0]{http://dx.doi.org/}%
\providecommand \selectlanguage [0]{\@gobble}%
\providecommand \bibinfo  [0]{\@secondoftwo}%
\providecommand \bibfield  [0]{\@secondoftwo}%
\providecommand \translation [1]{[#1]}%
\providecommand \BibitemOpen [0]{}%
\providecommand \bibitemStop [0]{}%
\providecommand \bibitemNoStop [0]{.\EOS\space}%
\providecommand \EOS [0]{\spacefactor3000\relax}%
\providecommand \BibitemShut  [1]{\csname bibitem#1\endcsname}%
\let\auto@bib@innerbib\@empty
%</preamble>
\bibitem [{\citenamefont {Nielsen}\ and\ \citenamefont
  {Chuang}(2010)}]{Nielsen:2010}%
  \BibitemOpen
  \bibfield  {author} {\bibinfo {author} {\bibfnamefont {M.~A.}\ \bibnamefont
  {Nielsen}}\ and\ \bibinfo {author} {\bibfnamefont {I.~L.}\ \bibnamefont
  {Chuang}},\ }\href {\doibase 10.1017/CBO9780511976667} {\emph {\bibinfo
  {title} {Quantum Computation and Quantum Information: 10th Anniversary
  Edition}}}\ (\bibinfo  {publisher} {Cambridge University Press},\ \bibinfo
  {year} {2010})\BibitemShut {NoStop}%
\bibitem [{\citenamefont {Weiss}(2012)}]{Weiss:2012}%
  \BibitemOpen
  \bibfield  {author} {\bibinfo {author} {\bibfnamefont {U.}~\bibnamefont
  {Weiss}},\ }\href@noop {} {\emph {\bibinfo {title} {Quantum Dissipative
  Systems}}}\ (\bibinfo  {publisher} {World Scientific},\ \bibinfo {address}
  {Singapur},\ \bibinfo {year} {2012})\BibitemShut {NoStop}%
\bibitem [{\citenamefont {Breuer}\ and\ \citenamefont
  {Petruccione}(2002)}]{BRE02}%
  \BibitemOpen
  \bibfield  {author} {\bibinfo {author} {\bibfnamefont {H.~P.}\ \bibnamefont
  {Breuer}}\ and\ \bibinfo {author} {\bibfnamefont {F.}~\bibnamefont
  {Petruccione}},\ }\href@noop {} {\emph {\bibinfo {title} {The theory of open
  quantum systems}}}\ (\bibinfo  {publisher} {Oxford University Press},\
  \bibinfo {address} {Great Clarendon Street},\ \bibinfo {year}
  {2002})\BibitemShut {NoStop}%
\bibitem [{\citenamefont {Verstraete}\ \emph {et~al.}(2009)\citenamefont
  {Verstraete}, \citenamefont {Wolf},\ and\ \citenamefont
  {Ignacio~Cirac}}]{Verstaete:2009}%
  \BibitemOpen
  \bibfield  {author} {\bibinfo {author} {\bibfnamefont {F.}~\bibnamefont
  {Verstraete}}, \bibinfo {author} {\bibfnamefont {M.~M.}\ \bibnamefont
  {Wolf}}, \ and\ \bibinfo {author} {\bibfnamefont {J.}~\bibnamefont
  {Ignacio~Cirac}},\ }\bibfield  {title} {\emph {\bibinfo {title} {Quantum
  computation and quantum-state engineering driven by dissipation},}\ }\href
  {https://doi.org/10.1038/nphys1342} {\bibfield  {journal} {\bibinfo
  {journal} {Nat. Phys.}\ }\textbf {\bibinfo {volume} {5}},\ \bibinfo {pages}
  {633} (\bibinfo {year} {2009})}\ \BibitemShut {NoStop}%
\bibitem [{\citenamefont {Mirrahimi}\ \emph {et~al.}(2014)\citenamefont
  {Mirrahimi}, \citenamefont {Leghtas}, \citenamefont {Albert}, \citenamefont
  {Touzard}, \citenamefont {Schoelkopf}, \citenamefont {Jiang},\ and\
  \citenamefont {Devoret}}]{Mirrahimi:2014}%
  \BibitemOpen
  \bibfield  {author} {\bibinfo {author} {\bibfnamefont {M.}~\bibnamefont
  {Mirrahimi}}, \bibinfo {author} {\bibfnamefont {Z.}~\bibnamefont {Leghtas}},
  \bibinfo {author} {\bibfnamefont {V.~V.}\ \bibnamefont {Albert}}, \bibinfo
  {author} {\bibfnamefont {S.}~\bibnamefont {Touzard}}, \bibinfo {author}
  {\bibfnamefont {R.~J.}\ \bibnamefont {Schoelkopf}}, \bibinfo {author}
  {\bibfnamefont {L.}~\bibnamefont {Jiang}}, \ and\ \bibinfo {author}
  {\bibfnamefont {M.~H.}\ \bibnamefont {Devoret}},\ }\bibfield  {title} {\emph
  {\bibinfo {title} {Dynamically protected cat-qubits: a new paradigm for
  universal quantum computation},}\ }\href {\doibase
  10.1088/1367-2630/16/4/045014} {\bibfield  {journal} {\bibinfo  {journal}
  {New J. Phys}\ }\textbf {\bibinfo {volume} {16}},\ \bibinfo {pages} {045014}
  (\bibinfo {year} {2014})}\ \BibitemShut {NoStop}%
\bibitem [{\citenamefont {Poyatos}\ \emph {et~al.}(1996)\citenamefont
  {Poyatos}, \citenamefont {Cirac},\ and\ \citenamefont
  {Zoller}}]{Poyatos:1996}%
  \BibitemOpen
  \bibfield  {author} {\bibinfo {author} {\bibfnamefont {J.~F.}\ \bibnamefont
  {Poyatos}}, \bibinfo {author} {\bibfnamefont {J.~I.}\ \bibnamefont {Cirac}},
  \ and\ \bibinfo {author} {\bibfnamefont {P.}~\bibnamefont {Zoller}},\
  }\bibfield  {title} {\emph {\bibinfo {title} {Quantum Reservoir Engineering
  with Laser Cooled Trapped Ions},}\ }\href {\doibase
  10.1103/PhysRevLett.77.4728} {\bibfield  {journal} {\bibinfo  {journal}
  {Phys. Rev. Lett.}\ }\textbf {\bibinfo {volume} {77}},\ \bibinfo {pages}
  {4728} (\bibinfo {year} {1996})}\ \BibitemShut {NoStop}%
\bibitem [{\citenamefont {Shankar}\ \emph {et~al.}(2013)\citenamefont
  {Shankar}, \citenamefont {Hatridge}, \citenamefont {Leghtas}, \citenamefont
  {Sliwa}, \citenamefont {Narla}, \citenamefont {Vool}, \citenamefont {Girvin},
  \citenamefont {Frunzio}, \citenamefont {Mirrahimi},\ and\ \citenamefont
  {Devoret}}]{Shankar:2013}%
  \BibitemOpen
  \bibfield  {author} {\bibinfo {author} {\bibfnamefont {S.}~\bibnamefont
  {Shankar}}, \bibinfo {author} {\bibfnamefont {M.}~\bibnamefont {Hatridge}},
  \bibinfo {author} {\bibfnamefont {Z.}~\bibnamefont {Leghtas}}, \bibinfo
  {author} {\bibfnamefont {K.~M.}\ \bibnamefont {Sliwa}}, \bibinfo {author}
  {\bibfnamefont {A.}~\bibnamefont {Narla}}, \bibinfo {author} {\bibfnamefont
  {U.}~\bibnamefont {Vool}}, \bibinfo {author} {\bibfnamefont {S.~M.}\
  \bibnamefont {Girvin}}, \bibinfo {author} {\bibfnamefont {L.}~\bibnamefont
  {Frunzio}}, \bibinfo {author} {\bibfnamefont {M.}~\bibnamefont {Mirrahimi}},
  \ and\ \bibinfo {author} {\bibfnamefont {M.~H.}\ \bibnamefont {Devoret}},\
  }\bibfield  {title} {\emph {\bibinfo {title} {Autonomously stabilized
  entanglement between two superconducting quantum bits},}\ }\href
  {https://doi.org/10.1038/nature12802} {\bibfield  {journal} {\bibinfo
  {journal} {Nature}\ }\textbf {\bibinfo {volume} {504}},\ \bibinfo {pages}
  {419} (\bibinfo {year} {2013})}\ \BibitemShut {NoStop}%
\bibitem [{\citenamefont {Tuorila}\ \emph {et~al.}(2019)\citenamefont
  {Tuorila}, \citenamefont {Stockburger}, \citenamefont {Ala-Nissila},
  \citenamefont {Ankerhold},\ and\ \citenamefont
  {M\"ott\"onen}}]{Ankerhold:2019}%
  \BibitemOpen
  \bibfield  {author} {\bibinfo {author} {\bibfnamefont {J.}~\bibnamefont
  {Tuorila}}, \bibinfo {author} {\bibfnamefont {J.}~\bibnamefont
  {Stockburger}}, \bibinfo {author} {\bibfnamefont {T.}~\bibnamefont
  {Ala-Nissila}}, \bibinfo {author} {\bibfnamefont {J.}~\bibnamefont
  {Ankerhold}}, \ and\ \bibinfo {author} {\bibfnamefont {M.}~\bibnamefont
  {M\"ott\"onen}},\ }\bibfield  {title} {\emph {\bibinfo {title}
  {System-environment correlations in qubit initialization and control},}\
  }\href {\doibase 10.1103/PhysRevResearch.1.013004} {\bibfield  {journal}
  {\bibinfo  {journal} {Phys. Rev. Research}\ }\textbf {\bibinfo {volume}
  {1}},\ \bibinfo {pages} {013004} (\bibinfo {year} {2019})}\ \BibitemShut
  {NoStop}%
\bibitem [{\citenamefont {Caldeira}\ and\ \citenamefont
  {Leggett}(1981)}]{Caldeira:1981tun}%
  \BibitemOpen
  \bibfield  {author} {\bibinfo {author} {\bibfnamefont {A.~O.}\ \bibnamefont
  {Caldeira}}\ and\ \bibinfo {author} {\bibfnamefont {A.~J.}\ \bibnamefont
  {Leggett}},\ }\bibfield  {title} {\emph {\bibinfo {title} {Influence of
  Dissipation on Quantum Tunneling in Macroscopic Systems},}\ }\href {\doibase
  10.1103/PhysRevLett.46.211} {\bibfield  {journal} {\bibinfo  {journal} {Phys.
  Rev. Lett.}\ }\textbf {\bibinfo {volume} {46}},\ \bibinfo {pages} {211}
  (\bibinfo {year} {1981})}\ \BibitemShut {NoStop}%
\bibitem [{\citenamefont {Caldeira}\ and\ \citenamefont
  {Leggett}(1983)}]{Caldeira:1982ann}%
  \BibitemOpen
  \bibfield  {author} {\bibinfo {author} {\bibfnamefont {A.}~\bibnamefont
  {Caldeira}}\ and\ \bibinfo {author} {\bibfnamefont {A.}~\bibnamefont
  {Leggett}},\ }\bibfield  {title} {\emph {\bibinfo {title} {Quantum tunnelling
  in a dissipative system},}\ }\href {\doibase
  https://doi.org/10.1016/0003-4916(83)90202-6} {\bibfield  {journal} {\bibinfo
   {journal} {Ann. Phys.}\ }\textbf {\bibinfo {volume} {149}},\ \bibinfo
  {pages} {374 } (\bibinfo {year} {1983})}\ \BibitemShut {NoStop}%
\bibitem [{\citenamefont {Leggett}\ \emph {et~al.}(1987)\citenamefont
  {Leggett}, \citenamefont {Chakravarty}, \citenamefont {Dorsey}, \citenamefont
  {Fisher}, \citenamefont {Garg},\ and\ \citenamefont
  {Zwerger}}]{Leggett:1987}%
  \BibitemOpen
  \bibfield  {author} {\bibinfo {author} {\bibfnamefont {A.~J.}\ \bibnamefont
  {Leggett}}, \bibinfo {author} {\bibfnamefont {S.}~\bibnamefont
  {Chakravarty}}, \bibinfo {author} {\bibfnamefont {A.~T.}\ \bibnamefont
  {Dorsey}}, \bibinfo {author} {\bibfnamefont {M.~P.~A.}\ \bibnamefont
  {Fisher}}, \bibinfo {author} {\bibfnamefont {A.}~\bibnamefont {Garg}}, \ and\
  \bibinfo {author} {\bibfnamefont {W.}~\bibnamefont {Zwerger}},\ }\bibfield
  {title} {\emph {\bibinfo {title} {Dynamics of the dissipative two-state
  system},}\ }\href {\doibase 10.1103/RevModPhys.59.1} {\bibfield  {journal}
  {\bibinfo  {journal} {Rev. Mod. Phys.}\ }\textbf {\bibinfo {volume} {59}},\
  \bibinfo {pages} {1} (\bibinfo {year} {1987})}\ \BibitemShut {NoStop}%
\bibitem [{\citenamefont {Dorsey}\ \emph {et~al.}(1986)\citenamefont {Dorsey},
  \citenamefont {Fisher},\ and\ \citenamefont {Wartak}}]{Dorsey:1986}%
  \BibitemOpen
  \bibfield  {author} {\bibinfo {author} {\bibfnamefont {A.~T.}\ \bibnamefont
  {Dorsey}}, \bibinfo {author} {\bibfnamefont {M.~P.~A.}\ \bibnamefont
  {Fisher}}, \ and\ \bibinfo {author} {\bibfnamefont {M.~S.}\ \bibnamefont
  {Wartak}},\ }\bibfield  {title} {\emph {\bibinfo {title} {Truncation scheme
  for double-well systems with Ohmic dissipation},}\ }\href {\doibase
  10.1103/PhysRevA.33.1117} {\bibfield  {journal} {\bibinfo  {journal} {Phys.
  Rev. A}\ }\textbf {\bibinfo {volume} {33}},\ \bibinfo {pages} {1117}
  (\bibinfo {year} {1986})}\ \BibitemShut {NoStop}%
\bibitem [{\citenamefont {Bray}\ and\ \citenamefont {Moore}(1982)}]{Bray:1982}%
  \BibitemOpen
  \bibfield  {author} {\bibinfo {author} {\bibfnamefont {A.~J.}\ \bibnamefont
  {Bray}}\ and\ \bibinfo {author} {\bibfnamefont {M.~A.}\ \bibnamefont
  {Moore}},\ }\bibfield  {title} {\emph {\bibinfo {title} {Influence of
  Dissipation on Quantum Coherence},}\ }\href {\doibase
  10.1103/PhysRevLett.49.1545} {\bibfield  {journal} {\bibinfo  {journal}
  {Phys. Rev. Lett.}\ }\textbf {\bibinfo {volume} {49}},\ \bibinfo {pages}
  {1545} (\bibinfo {year} {1982})}\ \BibitemShut {NoStop}%
\bibitem [{\citenamefont {Chakravarty}(1982)}]{Chakravarty:1982}%
  \BibitemOpen
  \bibfield  {author} {\bibinfo {author} {\bibfnamefont {S.}~\bibnamefont
  {Chakravarty}},\ }\bibfield  {title} {\emph {\bibinfo {title} {Quantum
  Fluctuations in the Tunneling between Superconductors},}\ }\href {\doibase
  10.1103/PhysRevLett.49.681} {\bibfield  {journal} {\bibinfo  {journal} {Phys.
  Rev. Lett.}\ }\textbf {\bibinfo {volume} {49}},\ \bibinfo {pages} {681}
  (\bibinfo {year} {1982})}\ \BibitemShut {NoStop}%
\bibitem [{\citenamefont {Weiss}\ \emph {et~al.}(1987)\citenamefont {Weiss},
  \citenamefont {Grabert}, \citenamefont {H\"anggi},\ and\ \citenamefont
  {Riseborough}}]{Grabert:1987}%
  \BibitemOpen
  \bibfield  {author} {\bibinfo {author} {\bibfnamefont {U.}~\bibnamefont
  {Weiss}}, \bibinfo {author} {\bibfnamefont {H.}~\bibnamefont {Grabert}},
  \bibinfo {author} {\bibfnamefont {P.}~\bibnamefont {H\"anggi}}, \ and\
  \bibinfo {author} {\bibfnamefont {P.}~\bibnamefont {Riseborough}},\
  }\bibfield  {title} {\emph {\bibinfo {title} {Incoherent tunneling in a
  double well},}\ }\href {\doibase 10.1103/PhysRevB.35.9535} {\bibfield
  {journal} {\bibinfo  {journal} {Phys. Rev. B}\ }\textbf {\bibinfo {volume}
  {35}},\ \bibinfo {pages} {9535} (\bibinfo {year} {1987})}\ \BibitemShut
  {NoStop}%
\bibitem [{\citenamefont {Matsuo}\ \emph {et~al.}(2006)\citenamefont {Matsuo},
  \citenamefont {Natsume},\ and\ \citenamefont {Kato}}]{Matsuo1:2004}%
  \BibitemOpen
  \bibfield  {author} {\bibinfo {author} {\bibfnamefont {T.}~\bibnamefont
  {Matsuo}}, \bibinfo {author} {\bibfnamefont {Y.}~\bibnamefont {Natsume}}, \
  and\ \bibinfo {author} {\bibfnamefont {T.}~\bibnamefont {Kato}},\ }\bibfield
  {title} {\emph {\bibinfo {title} {Quantum-Classical Transition in Dissipative
  Double-Well Systems â A Numerical Study by a New Monte Carlo Scheme
  â},}\ }\href {\doibase 10.1143/JPSJ.75.103002} {\bibfield  {journal}
  {\bibinfo  {journal} {Journal of the Physical Society of Japan}\ }\textbf
  {\bibinfo {volume} {75}},\ \bibinfo {pages} {103002} (\bibinfo {year}
  {2006})}\ \BibitemShut {NoStop}%
\bibitem [{\citenamefont {Matsuo}\ \emph {et~al.}(2008)\citenamefont {Matsuo},
  \citenamefont {Natsume},\ and\ \citenamefont {Kato}}]{Matsuo2:2008}%
  \BibitemOpen
  \bibfield  {author} {\bibinfo {author} {\bibfnamefont {T.}~\bibnamefont
  {Matsuo}}, \bibinfo {author} {\bibfnamefont {Y.}~\bibnamefont {Natsume}}, \
  and\ \bibinfo {author} {\bibfnamefont {T.}~\bibnamefont {Kato}},\ }\bibfield
  {title} {\emph {\bibinfo {title} {Quantum-classical transition and
  decoherence in dissipative double-well potential systems: Monte Carlo
  algorithm},}\ }\href {\doibase 10.1103/PhysRevB.77.184304} {\bibfield
  {journal} {\bibinfo  {journal} {Phys. Rev. B}\ }\textbf {\bibinfo {volume}
  {77}},\ \bibinfo {pages} {184304} (\bibinfo {year} {2008})}\ \BibitemShut
  {NoStop}%
\bibitem [{\citenamefont {Aoki}\ and\ \citenamefont
  {Horikoshi}(2002)}]{Aoki:2002}%
  \BibitemOpen
  \bibfield  {author} {\bibinfo {author} {\bibfnamefont {K.-I.}\ \bibnamefont
  {Aoki}}\ and\ \bibinfo {author} {\bibfnamefont {A.}~\bibnamefont
  {Horikoshi}},\ }\bibfield  {title} {\emph {\bibinfo {title} {Nonperturbative
  renormalization-group approach for quantum dissipative systems},}\ }\href
  {\doibase 10.1103/PhysRevA.66.042105} {\bibfield  {journal} {\bibinfo
  {journal} {Phys. Rev. A}\ }\textbf {\bibinfo {volume} {66}},\ \bibinfo
  {pages} {042105} (\bibinfo {year} {2002})}\ \BibitemShut {NoStop}%
\bibitem [{\citenamefont {Kov{\'a}cs}\ \emph {et~al.}(2017)\citenamefont
  {Kov{\'a}cs}, \citenamefont {Fazekas}, \citenamefont {Nagy},\ and\
  \citenamefont {Sailer}}]{Kovacs:2017}%
  \BibitemOpen
  \bibfield  {author} {\bibinfo {author} {\bibfnamefont {J.}~\bibnamefont
  {Kov{\'a}cs}}, \bibinfo {author} {\bibfnamefont {B.}~\bibnamefont {Fazekas}},
  \bibinfo {author} {\bibfnamefont {S.}~\bibnamefont {Nagy}}, \ and\ \bibinfo
  {author} {\bibfnamefont {K.}~\bibnamefont {Sailer}},\ }\bibfield  {title}
  {{\selectlanguage {English}\emph {\bibinfo {title} {Quantum classical
  transition in the Caldeira Leggett model},}\ }}\href {\doibase
  10.1016/j.aop.2016.12.010} {\bibfield  {journal} {\bibinfo  {journal} {Ann.
  Phys.}\ }\textbf {\bibinfo {volume} {376}},\ \bibinfo {pages} {372} (\bibinfo
  {year} {2017})}\ \BibitemShut {NoStop}%
\bibitem [{\citenamefont {Fujikawa}\ \emph {et~al.}(1992)\citenamefont
  {Fujikawa}, \citenamefont {Iso}, \citenamefont {Sasaki},\ and\ \citenamefont
  {Suzuki}}]{Fujikawa:1992}%
  \BibitemOpen
  \bibfield  {author} {\bibinfo {author} {\bibfnamefont {K.}~\bibnamefont
  {Fujikawa}}, \bibinfo {author} {\bibfnamefont {S.}~\bibnamefont {Iso}},
  \bibinfo {author} {\bibfnamefont {M.}~\bibnamefont {Sasaki}}, \ and\ \bibinfo
  {author} {\bibfnamefont {H.}~\bibnamefont {Suzuki}},\ }\bibfield  {title}
  {\emph {\bibinfo {title} {Quantum tunneling with dissipation: Possible
  enhancement by dissipative interactions},}\ }\href {\doibase
  10.1103/PhysRevB.46.10295} {\bibfield  {journal} {\bibinfo  {journal} {Phys.
  Rev. B}\ }\textbf {\bibinfo {volume} {46}},\ \bibinfo {pages} {10295}
  (\bibinfo {year} {1992})}\ \BibitemShut {NoStop}%
\bibitem [{\citenamefont {Ankerhold}\ and\ \citenamefont
  {Pollak}(2007)}]{Ankerhold:2007}%
  \BibitemOpen
  \bibfield  {author} {\bibinfo {author} {\bibfnamefont {J.}~\bibnamefont
  {Ankerhold}}\ and\ \bibinfo {author} {\bibfnamefont {E.}~\bibnamefont
  {Pollak}},\ }\bibfield  {title} {\emph {\bibinfo {title} {Dissipation can
  enhance quantum effects},}\ }\href {\doibase 10.1103/PhysRevE.75.041103}
  {\bibfield  {journal} {\bibinfo  {journal} {Phys. Rev. E}\ }\textbf {\bibinfo
  {volume} {75}},\ \bibinfo {pages} {041103} (\bibinfo {year} {2007})}\
  \BibitemShut {NoStop}%
\bibitem [{\citenamefont {Cuccoli}\ \emph {et~al.}(2010)\citenamefont
  {Cuccoli}, \citenamefont {Del~Sette},\ and\ \citenamefont
  {Vaia}}]{Cuccoli:2010dr}%
  \BibitemOpen
  \bibfield  {author} {\bibinfo {author} {\bibfnamefont {A.}~\bibnamefont
  {Cuccoli}}, \bibinfo {author} {\bibfnamefont {N.}~\bibnamefont {Del~Sette}},
  \ and\ \bibinfo {author} {\bibfnamefont {R.}~\bibnamefont {Vaia}},\
  }\bibfield  {title} {\emph {\bibinfo {title} {{Reentrant enhancement of
  quantum fluctuations for symmetric environmental coupling}},}\ }\href@noop {}
  {\bibfield  {journal} {\bibinfo  {journal} {Phys. Rev. E}\ }\textbf {\bibinfo
  {volume} {81}},\ \bibinfo {pages} {041110} (\bibinfo {year} {2010})}\
  \BibitemShut {NoStop}%
\bibitem [{\citenamefont {Rastelli}(2016)}]{Rastelli:2016ge}%
  \BibitemOpen
  \bibfield  {author} {\bibinfo {author} {\bibfnamefont {G.}~\bibnamefont
  {Rastelli}},\ }\bibfield  {title} {\emph {\bibinfo {title}
  {{Dissipation-induced enhancement of quantum fluctuations}},}\ }\href@noop {}
  {\bibfield  {journal} {\bibinfo  {journal} {New Journal of Physics}\ }\textbf
  {\bibinfo {volume} {18}},\ \bibinfo {pages} {053033} (\bibinfo {year}
  {2016})}\ \BibitemShut {NoStop}%
\bibitem [{\citenamefont {Kohler}\ and\ \citenamefont
  {Sols}(2006)}]{Kohler:2006ky}%
  \BibitemOpen
  \bibfield  {author} {\bibinfo {author} {\bibfnamefont {H.}~\bibnamefont
  {Kohler}}\ and\ \bibinfo {author} {\bibfnamefont {F.}~\bibnamefont {Sols}},\
  }\bibfield  {title} {\emph {\bibinfo {title} {{Dissipative quantum oscillator
  with two competing heat baths}},}\ }\href@noop {} {\bibfield  {journal}
  {\bibinfo  {journal} {New Journal of Physics}\ }\textbf {\bibinfo {volume}
  {8}},\ \bibinfo {pages} {149} (\bibinfo {year} {2006})}\ \BibitemShut
  {NoStop}%
\bibitem [{\citenamefont {Devoret}(2002)}]{Devoret:2004-LesHouches}%
  \BibitemOpen
  \bibfield  {author} {\bibinfo {author} {\bibfnamefont {M.~H.}\ \bibnamefont
  {Devoret}},\ }in\ \href@noop {} {\emph {\bibinfo {booktitle} {Quantum
  Fluctuations (Les Houches Session LXIII)}}}\ (\bibinfo  {publisher}
  {Elsevier},\ \bibinfo {address} {Amsterdam},\ \bibinfo {year} {2002})\
  p.~\bibinfo {pages} {1}\BibitemShut {NoStop}%
\bibitem [{\citenamefont {Fazio}\ and\ \citenamefont {van~der
  Zant}(2001)}]{Fazio:2001}%
  \BibitemOpen
  \bibfield  {author} {\bibinfo {author} {\bibfnamefont {R.}~\bibnamefont
  {Fazio}}\ and\ \bibinfo {author} {\bibfnamefont {H.}~\bibnamefont {van~der
  Zant}},\ }\bibfield  {title} {\emph {\bibinfo {title} {Quantum phase
  transitions and vortex dynamics in superconducting networks},}\ }\href
  {\doibase http://dx.doi.org/10.1016/S0370-1573(01)00022-9} {\bibfield
  {journal} {\bibinfo  {journal} {Physics Reports}\ }\textbf {\bibinfo {volume}
  {355}},\ \bibinfo {pages} {235 } (\bibinfo {year} {2001})}\ \BibitemShut
  {NoStop}%
\bibitem [{\citenamefont {Liu}\ \emph {et~al.}(2014)\citenamefont {Liu},
  \citenamefont {Zheng}, \citenamefont {Finkelstein},\ and\ \citenamefont
  {Baranger}}]{Baranger:2014}%
  \BibitemOpen
  \bibfield  {author} {\bibinfo {author} {\bibfnamefont {D.~E.}\ \bibnamefont
  {Liu}}, \bibinfo {author} {\bibfnamefont {H.}~\bibnamefont {Zheng}}, \bibinfo
  {author} {\bibfnamefont {G.}~\bibnamefont {Finkelstein}}, \ and\ \bibinfo
  {author} {\bibfnamefont {H.~U.}\ \bibnamefont {Baranger}},\ }\bibfield
  {title} {\emph {\bibinfo {title} {Tunable quantum phase transitions in a
  resonant level coupled to two dissipative baths},}\ }\href {\doibase
  10.1103/PhysRevB.89.085116} {\bibfield  {journal} {\bibinfo  {journal} {Phys.
  Rev. B}\ }\textbf {\bibinfo {volume} {89}},\ \bibinfo {pages} {085116}
  (\bibinfo {year} {2014})}\ \BibitemShut {NoStop}%
\bibitem [{\citenamefont {Otten}\ and\ \citenamefont
  {Hassler}(2017)}]{Otten:2017}%
  \BibitemOpen
  \bibfield  {author} {\bibinfo {author} {\bibfnamefont {D.}~\bibnamefont
  {Otten}}\ and\ \bibinfo {author} {\bibfnamefont {F.}~\bibnamefont
  {Hassler}},\ }\bibfield  {title} {\emph {\bibinfo {title} {Korshunov
  instantons in a superconductor at elevated bias current},}\ }\href {\doibase
  10.1103/PhysRevB.96.125433} {\bibfield  {journal} {\bibinfo  {journal} {Phys.
  Rev. B}\ }\textbf {\bibinfo {volume} {96}},\ \bibinfo {pages} {125433}
  (\bibinfo {year} {2017})}\ \BibitemShut {NoStop}%
\bibitem [{\citenamefont {Maile}\ \emph {et~al.}(2018)\citenamefont {Maile},
  \citenamefont {Andergassen}, \citenamefont {Belzig},\ and\ \citenamefont
  {Rastelli}}]{Maile:2018}%
  \BibitemOpen
  \bibfield  {author} {\bibinfo {author} {\bibfnamefont {D.}~\bibnamefont
  {Maile}}, \bibinfo {author} {\bibfnamefont {S.}~\bibnamefont {Andergassen}},
  \bibinfo {author} {\bibfnamefont {W.}~\bibnamefont {Belzig}}, \ and\ \bibinfo
  {author} {\bibfnamefont {G.}~\bibnamefont {Rastelli}},\ }\bibfield  {title}
  {\emph {\bibinfo {title} {Quantum phase transition with dissipative
  frustration},}\ }\href {\doibase 10.1103/PhysRevB.97.155427} {\bibfield
  {journal} {\bibinfo  {journal} {Phys. Rev. B}\ }\textbf {\bibinfo {volume}
  {97}},\ \bibinfo {pages} {155427} (\bibinfo {year} {2018})}\ \BibitemShut
  {NoStop}%
\bibitem [{\citenamefont {Bruognolo}\ \emph {et~al.}(2014)\citenamefont
  {Bruognolo}, \citenamefont {Weichselbaum}, \citenamefont {Guo}, \citenamefont
  {von Delft}, \citenamefont {Schneider},\ and\ \citenamefont
  {Vojta}}]{Bruognolo:2014jf}%
  \BibitemOpen
  \bibfield  {author} {\bibinfo {author} {\bibfnamefont {B.}~\bibnamefont
  {Bruognolo}}, \bibinfo {author} {\bibfnamefont {A.}~\bibnamefont
  {Weichselbaum}}, \bibinfo {author} {\bibfnamefont {C.}~\bibnamefont {Guo}},
  \bibinfo {author} {\bibfnamefont {J.}~\bibnamefont {von Delft}}, \bibinfo
  {author} {\bibfnamefont {I.}~\bibnamefont {Schneider}}, \ and\ \bibinfo
  {author} {\bibfnamefont {M.}~\bibnamefont {Vojta}},\ }\bibfield  {title}
  {\emph {\bibinfo {title} {{Two-bath spin-boson model: Phase diagram and
  critical properties}},}\ }\href@noop {} {\bibfield  {journal} {\bibinfo
  {journal} {Phys. Rev. B}\ }\textbf {\bibinfo {volume} {90}},\ \bibinfo
  {pages} {245130} (\bibinfo {year} {2014})}\ \BibitemShut {NoStop}%
\bibitem [{\citenamefont {Castro~Neto}\ \emph {et~al.}(2003)\citenamefont
  {Castro~Neto}, \citenamefont {Novais}, \citenamefont {Borda}, \citenamefont
  {Zarand},\ and\ \citenamefont {Affleck}}]{Neto:2003ir}%
  \BibitemOpen
  \bibfield  {author} {\bibinfo {author} {\bibfnamefont {A.~H.}\ \bibnamefont
  {Castro~Neto}}, \bibinfo {author} {\bibfnamefont {E.}~\bibnamefont {Novais}},
  \bibinfo {author} {\bibfnamefont {L.}~\bibnamefont {Borda}}, \bibinfo
  {author} {\bibfnamefont {G.}~\bibnamefont {Zarand}}, \ and\ \bibinfo {author}
  {\bibfnamefont {I.}~\bibnamefont {Affleck}},\ }\bibfield  {title} {\emph
  {\bibinfo {title} {{Quantum Magnetic Impurities in Magnetically Ordered
  Systems}},}\ }\href@noop {} {\bibfield  {journal} {\bibinfo  {journal} {Phys.
  Rev. Lett}\ }\textbf {\bibinfo {volume} {91}},\ \bibinfo {pages} {096401}
  (\bibinfo {year} {2003})}\ \BibitemShut {NoStop}%
\bibitem [{\citenamefont {Coleman}(1979)}]{Coleman:1978ae}%
  \BibitemOpen
  \bibfield  {author} {\bibinfo {author} {\bibfnamefont {S.~R.}\ \bibnamefont
  {Coleman}},\ }\bibfield  {title} {\emph {\bibinfo {title} {{The Uses of
  Instantons}},}\ }\bibfield  {booktitle} {\emph {\bibinfo {booktitle}
  {{Instantons in gauge theories}}},\ }\href@noop {} {\bibfield  {journal}
  {\bibinfo  {journal} {Subnucl. Ser.}\ }\textbf {\bibinfo {volume} {15}},\
  \bibinfo {pages} {805} (\bibinfo {year} {1979})}\ \BibitemShut {NoStop}%
%%CITATION = SUSEE,15,805;%%
\bibitem [{\citenamefont {Kleinert}(2009)}]{Kleinert:2009}%
  \BibitemOpen
  \bibfield  {author} {\bibinfo {author} {\bibfnamefont {H.}~\bibnamefont
  {Kleinert}},\ }\href@noop {} {\emph {\bibinfo {title} {Path Integrals in
  Quantum Mechanics, Statistics, Polymer Physics, and Financial Markets -}}}\
  (\bibinfo  {publisher} {World Scientific},\ \bibinfo {address} {Singapur},\
  \bibinfo {year} {2009})\BibitemShut {NoStop}%
\bibitem [{\citenamefont {Anderson}\ \emph {et~al.}(1970)\citenamefont
  {Anderson}, \citenamefont {Yuval},\ and\ \citenamefont
  {Hamann}}]{Anderson:1970}%
  \BibitemOpen
  \bibfield  {author} {\bibinfo {author} {\bibfnamefont {P.~W.}\ \bibnamefont
  {Anderson}}, \bibinfo {author} {\bibfnamefont {G.}~\bibnamefont {Yuval}}, \
  and\ \bibinfo {author} {\bibfnamefont {D.~R.}\ \bibnamefont {Hamann}},\
  }\bibfield  {title} {\emph {\bibinfo {title} {Exact Results in the Kondo
  Problem. II. Scaling Theory, Qualitatively Correct Solution, and Some New
  Results on One-Dimensional Classical Statistical Models},}\ }\href {\doibase
  10.1103/PhysRevB.1.4464} {\bibfield  {journal} {\bibinfo  {journal} {Phys.
  Rev. B}\ }\textbf {\bibinfo {volume} {1}},\ \bibinfo {pages} {4464} (\bibinfo
  {year} {1970})}\ \BibitemShut {NoStop}%
\bibitem [{Note1()}]{Note1}%
  \BibitemOpen
  \bibinfo {note} {Note that the action on the classical path is convergent and
  the effect of the high frequency cutoff function $f_c(|\omega _l|/\omega _c)$
  is negligible.}\BibitemShut {Stop}%
\bibitem [{Note2()}]{Note2}%
  \BibitemOpen
  \bibinfo {note} {We use $V_0/\hbar \omega _0=4$ in Fig.~\ref
  {Fig:phasechange2}d to illustrate of the change of the activity due to the
  unconventional coupling. The $V_0$ dependence of the result is plotted in
  Fig.~\ref {Fig:phasechange}a-b.}\BibitemShut {Stop}%
\end{thebibliography}%

\end{document}